\documentclass{article}

\usepackage{arxiv}

\usepackage[utf8]{inputenc} 
\usepackage[T1]{fontenc}    
\usepackage{hyperref}       
\usepackage{url}            
\usepackage{booktabs}       
\usepackage{amsfonts}       
\usepackage{nicefrac}       
\usepackage{microtype}      
\usepackage{lipsum}
\usepackage{algorithmic}
\usepackage{algorithm}
\usepackage{hyperref}
\hypersetup{hidelinks}
\usepackage{multirow}
\usepackage{tikz}
\usepackage{subfigure}
\usepackage{amssymb}
\usepackage{amsmath}
\usepackage{amsthm}
\title{Communication Optimization Strategies for Distributed Deep Neural Network Training: A Survey}

\author{
  Shuo Ouyang\\
  Department of Computer Science\\
  National University of Defense Technology\\
  \texttt{ouyangshuo@nudt.edu.cn} \\
   \And
 Dezun Dong\thanks{Corresponding Author} \\
  Department of Computer Science\\
  National University of Defense Technology\\
  \texttt{dong@nudt.edu.cn} \\
  \And
  Yemao Xu \\
  Department of Computer Science\\
  National University of Defense Technology\\
  \texttt{xuyemaovip@nudt.edu.cn} \\
  \And
  Liquan Xiao \\
  Department of Computer Science\\
  National University of Defense Technology\\
  \texttt{xiaoliquan@nudt.edu.cn} \\
}

\begin{document}
\maketitle

\begin{abstract}
Recent trends in high-performance computing and deep learning have led to the proliferation of studies on large-scale deep neural network training. However, the frequent communication requirements among computation nodes drastically slows the overall training speeds, which causes bottlenecks in distributed training, particularly in clusters with limited network bandwidths. To mitigate the drawbacks of distributed communications, researchers have proposed various optimization strategies. In this paper, we provide a comprehensive survey of communication strategies from both an algorithm viewpoint and a computer network perspective. Algorithm optimizations focus on reducing the communication volumes used in distributed training, while network optimizations focus on accelerating the communications between distributed devices. At the algorithm level, we describe how to reduce the number of communication rounds and transmitted bits per round. In addition, we elucidate how to overlap computation and communication. At the network level, we discuss the effects caused by network infrastructures, including logical communication schemes and network protocols. Finally, we extrapolate the potential future challenges and new research directions to accelerate communications for distributed deep neural network training.

\end{abstract}


\section{Introduction}
\label{sec:introduction}
Currently, we are in an unprecedented and historic era of deep learning research in which waves of deep neural networks have swept through several application domains, ranging from autonomous driving\cite{chen2015deepdriving}, computer vision\cite{chollet2017xception, zhang2020resnest}, and natural language processing\cite{cho2014learning,brown2020language} to recommendation systems\cite{covington2016deep}. The popularity of deep learning promoted the development of DNN architectures, including fully connected neural networks (FCNN), convolutional neural networks (CNN), recurrent neural networks (RNN) and its variants (LSTM\cite{hochreiter1997long}, GRU\cite{cho2014learning}). These neural networks have achieved state-of-the-art performance across various domain-specific tasks. Taking computer vision as an example, well-designed DNNs, such as GoogLeNet\cite{szegedy2015going} and ResNet-50\cite{he2016deep} trained on ImageNet\cite{russakovsky2015imagenet} dataset, have beaten humans on image classification tasks.

Regrading good performance, DNNs tend to be deeper and more sophisticated and are trained with the larger datasets. The rapid increase of data volumes and model sizes have resulted in vast amounts of computation, indicating that DNN training is time-consuming and can extends over several days or weeks. High-performance hardware, such as graphics processing units (GPU) \cite{lindholm2008nvidia} and tensor processing units (TPU) \cite{jouppi2017datacenter}, are applied to accelerate the training time. 

Beyond using high-performance hardware, paralleling and deploying DNN training tasks on multiple nodes (consisting of one or more machines) is another practical approach. Under these conditions, each node only executes part of an entire computation task. However, due to the frequent communication requirements for exchanging large amounts of data among the different computation nodes, communications overhead is a critical bottleneck in distributed training. With the growth of the cluster scale, communications overhead has increased explosively. Such a phenomenon considerably diminishes the advantage of parallel training, as a majority of the training time is spent on transferring data. When high-performance hardware accelerators are used, the proportion of time spent on communication increases further because they only decrease the computation overhead, whereas the communications overhead is unchanged.

\begin{figure*}[ht]
 \centering
 \includegraphics[width=\textwidth, trim=10 0 170 0,clip]{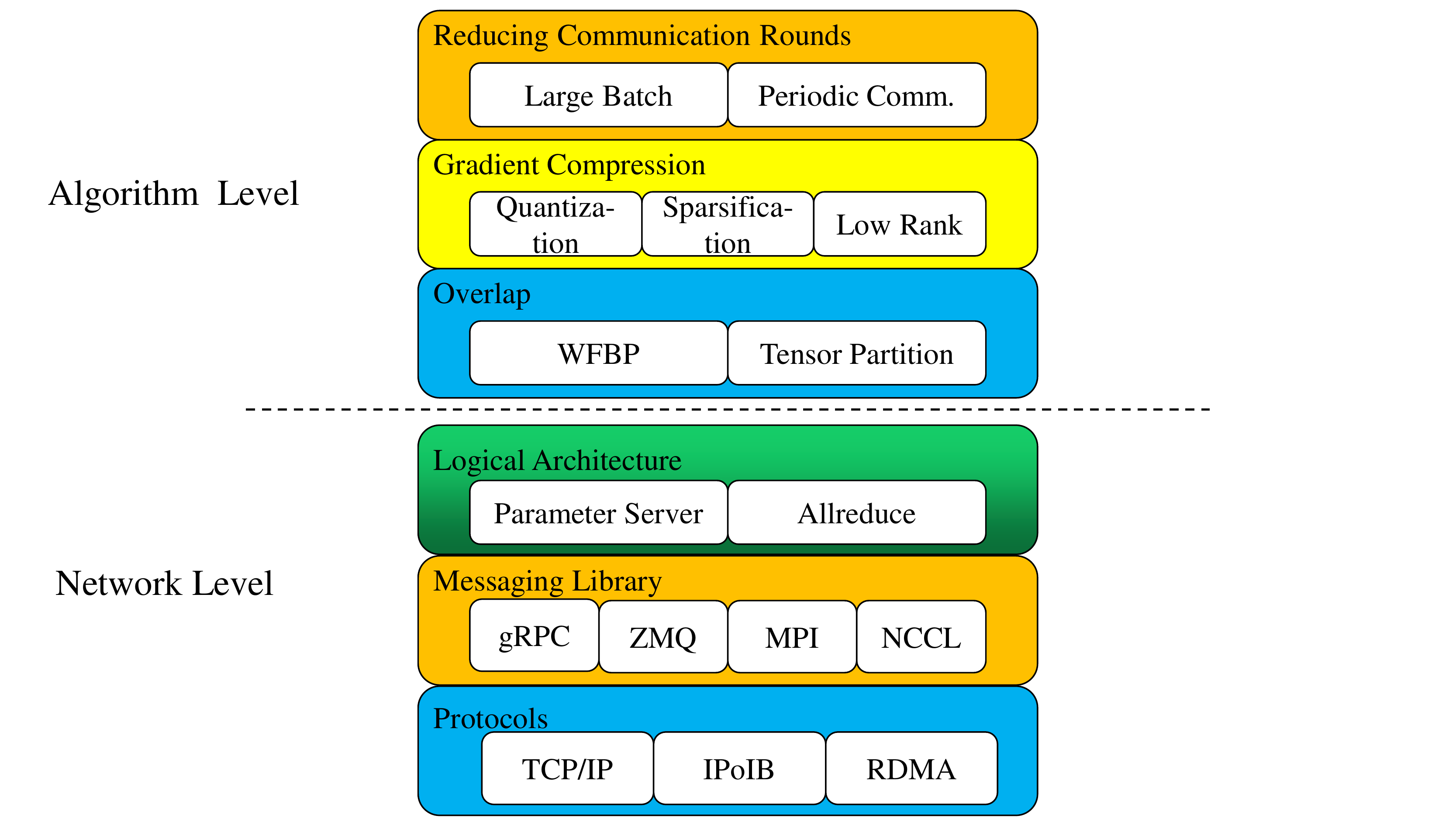}
 \caption{Overview of communication optimization strategies. Based on the characteristic of distributed training, we divide optimization methods into two categories: communication reduction and scheduling algorithm part (referred to as algorithm level) and network traffic execution part (referred to as network level) \label{fig:overview}}
\end{figure*}

In this paper, we primarily investigate how to deal with communications overhead for distributed DNN training. Because distributed deep learning is a cross-disciplinary field, both deep learning and distributed network communities have proposed communication optimization strategies from their own perspectives. In this survey, we bring together, classify, and compare the large body of work on communications optimization for distributed DNN training from the different research communities that contribute to this area. An overview of optimization strategies is shown in Fig.\ref{fig:overview}.

Previous studies have covered many research domains of distributed deep learning. Ben et al.\cite{ben2019demystifying} provided a detailed concurrence analysis of DNNs from different levels. Various training algorithms and modern state-of-the-art training frameworks were studied by Chahal et al.\cite{chahal2019hitchhiker}. A recent review by Mayer et al.\cite{mayer2020scalable} discusses the challenges of managing large deep learning systems on a distributed infrastructure. Mittal et al.\cite{mittal2019survey} reported optimization techniques for deep learning applications of GPUs from an  architecture and a system-level aspect. Some surveys have discussed communication optimization issues in distributed deep learning, but we provide a broader investigation. In particular, we discuss optimization from high algorithms level to low network level, which is a perspective that was missing from previous surveys.

The rest of this paper is organized as follows: In Section \ref{sec:background}, we introduce deep learning's background and provide an overview of distributed DNN training. In the next two sections, we discuss the optimization of algorithms and networks, respectively. Section \ref{sec:algorithm} discusses communication rounds reduction, gradient compression and computation-communication overlap, whereas Section \ref{sec:network} introduces logical communication architectures and network protocols. Finally, we conclude the entire paper and highlight potential challenges and research directions in Section \ref{sec:conclusion}.

\section{Background}
\label{sec:background}
In this section, we provide a brief introduction to large-scale distributed deep neural network training. Currently, the most popular method to train DNNs on a single-machine remains mini-batch based stochastic gradient descent (SGD) algorithm\cite{bottou2012stochastic, bottou2010large} with error back-propagation\cite{rumelhart1986learning}. When DNN training moves to parallelization, several problems need to be considered: (i) which part of the training task can be parallelized, (ii) the architectures of the computation nodes, and (iii) when to synchronize gradients.

\subsection{Stochastic Gradient Descent}

SGD and its variants have become the workhorse algorithms for training DNN models. Suppose that we want to train a DNN model to minimize the loss function $\ell(f(x; W), y)$ averaged on a given dataset $D$, where $f(x, W)$ represents the DNN's output with parameter $W$, and $x$ is an input data instance corresponding with the true label $y$. The application-specific loss function $\ell(\cdot)$ measures the difference between outputs and true labels. SGD will update model parameters and iteratively converge towards the minimum of the loss function as follows\cite{bottou2010large}:

$$
{W}_{t+1} = {W}_t - \eta \times \frac{1}{B} \sum_{i=1}^{B}\nabla \ell \left ( f \left (x_i; W_t \right), y_i \right)
$$

where $B$ is the number of training instance in each mini-batch, $\eta$ is learning rate and $\nabla \ell \left ( f \left (x_i; W_t \right), y_i \right)$ are the gradients of current mini-batch.

\subsection{Data and Model Parallelism}

\begin{figure*}[ht]
 \centering
 \subfigure[Data Parallelism]{
  \label{fig:data-parallel}
  \includegraphics[width=0.45\textwidth, trim=20 10 100 10, clip]{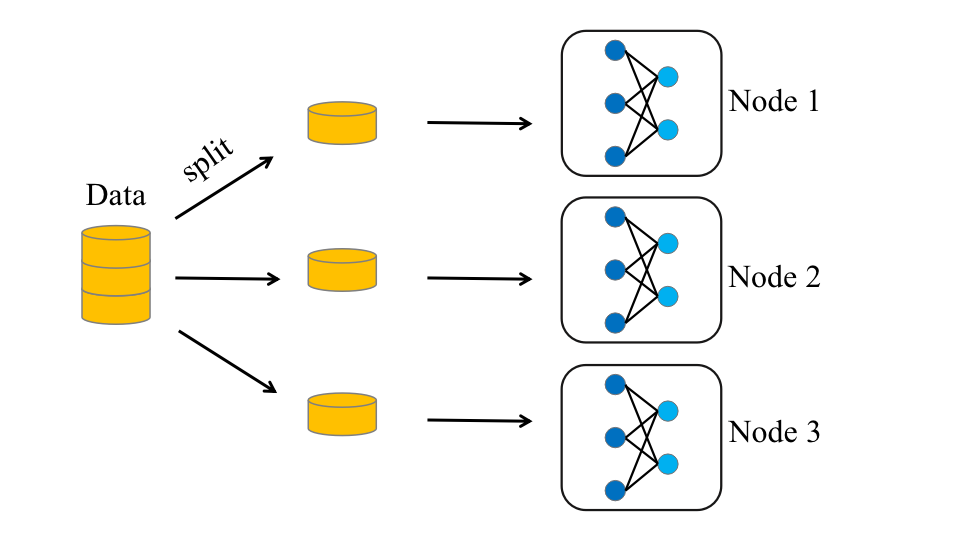}
 }
 \subfigure[Model Parallelism]{
  \label{fig:model-parallel}
  \includegraphics[width=0.45\textwidth, trim=40 10 40 10, clip]{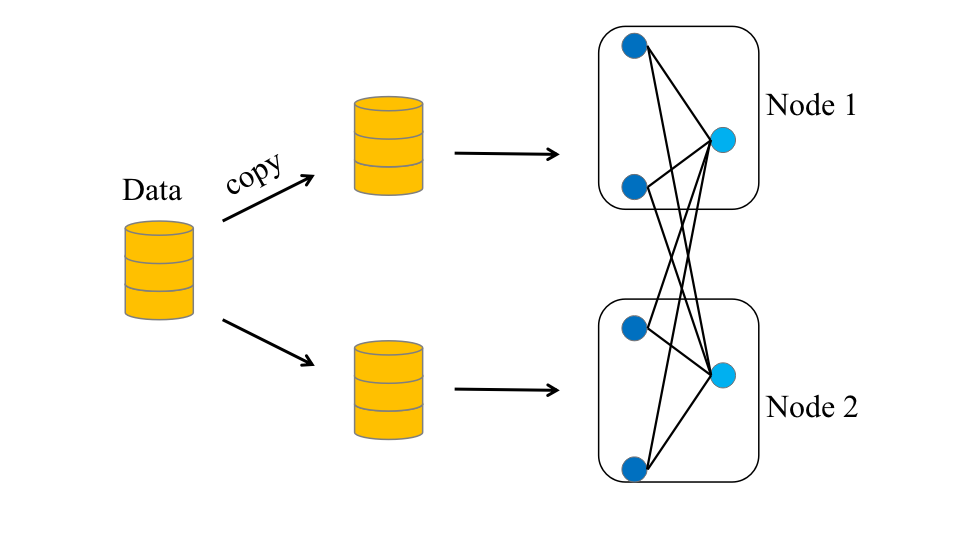}
 }
 \caption{DNN parallelism schemes.\label{fig:parallelism}}
\end{figure*}

Data parallelism and model parallelism are two commonly training methods used in distributed deep learning. In data parallelism, the entire training dataset is randomly and equally divided into $N$ parts and dispatched on $N$ nodes, see in Fig.\ref{fig:data-parallel}. Each node maintains a model replica along with its local parameters $W^{n}$. The training process is as follows:

\begin{enumerate}
 
 \item Each node reads a mini-batch of the training data and executes forward and backward propagations to calculate its local gradients $\nabla \ell \left ( f \left (x; W^{n}_{t} \right), y \right)$.
 
 \item Each node sends the local gradients to a master node. After receiving gradients from all the nodes, the master aggregates these gradients and updates the model by ${W}_{t+1} = {W}_{t} - \frac{\eta}{N}\sum^{N}_{n=1}\nabla \ell \left ( f \left (x; W^{n}_{t} \right), y \right)$.
 
 \item The master broadcasts the latest model parameters to all other nodes.
 
 \item Repeat these three steps until the model converges.
\end{enumerate}

With respect to model parallelism\cite{gholami2018integrated,dean2012large}, different parts of the DNN model are split into different nodes. Hence, the number of parameters on a single node is reduced. Nodes, where the input layer is located, are responsible for reading the entire dataset. For example, only Node 1 and Node 3 read input dataset in Fig. \ref{fig:model-parallel}. Correspondingly, nodes where the output layer is located are responsible for outputting the predicted value of the DNN. The calculations between nodes are no longer independent of each other. Only neurons with connections that cross computation nodes (thick line in Fig. \ref{fig:model-parallel}) will need to exchange gradients and model parameters.  Clearly, data parallelism is suitable for a large training dataset, while model parallelism is applicable when a model is too large to fit in a single node. In this paper, we primarily focus on data parallelism.

\subsection{Centralized and Dencentralized Architectures}

The (logic) architecture of the computation nodes can affect the communication modes and the network's performance. Parameter server\cite{li2014scaling, li2014communication} is the most popular centralized architecture in distributed deep learning. A parameter server usually includes a server node and several worker nodes. The server maintains global model parameters, whereas each worker stores a local model replica. If the server node has more than one machine, each machine maintains a partition of the entire model parameters. For workers, each worker stores an entire model replica in data parallelism or a part of the model in model parallelism. Workers communicate with the server via a \emph{push/pull} operation, whereas there is no communication between any workers. The drawback of the parameter server is the bandwidth bottleneck on the server-side with the increment of workers.

Due to this drawback of the parameter server, decentralized architectures have attracted considerable research attention because they incur fewer communication traffic issues\cite{lian2017can}. Similar to the parameter server, each node in a decentralized architecture holds an entire model replica, but they use collective communication operation \emph{Allreduce} instead of \emph{push/pull} to exchange gradients and parameters. Nevertheless, the Allreduce operation has a variety of implementations with notable differences in performance, which means that it may affect the communications overhead. We will discuss related issues in Section \ref{sec:network}.

\subsection{Synchronous and Asynchronous Updates}
Owing to differences in network bandwidth and computing power, some nodes may calculate gradients faster, while other nodes may be slower. The main challenge that comes with such circumstances is determining when to synchronize the gradients among multiple computation nodes. There are three different methods that reasonably solve this problem: synchronous, asynchronous, and bounded delay updates.

\subsubsection{Synchronous}
In the synchronous update, the server does not update the model until it receives gradients from all the workers at each iteration. In other words, faster workers will wait for slower workers. One well-known implementation of the synchronous update is bulk synchronous parallel (BSP)\cite{valiant1990bridging}. A characteristic of the synchronous mode is that the server will always receive the latest gradients of all the nodes, which does not affect the model's convergence. However, fast nodes idle when waiting for slow nodes, leading to a waste of resources and causing the straggler problem that slows the overall training time.

\subsubsection{Asynchronous}
\begin{figure*}[ht]
 \centering
 \includegraphics[width=0.8\textwidth]{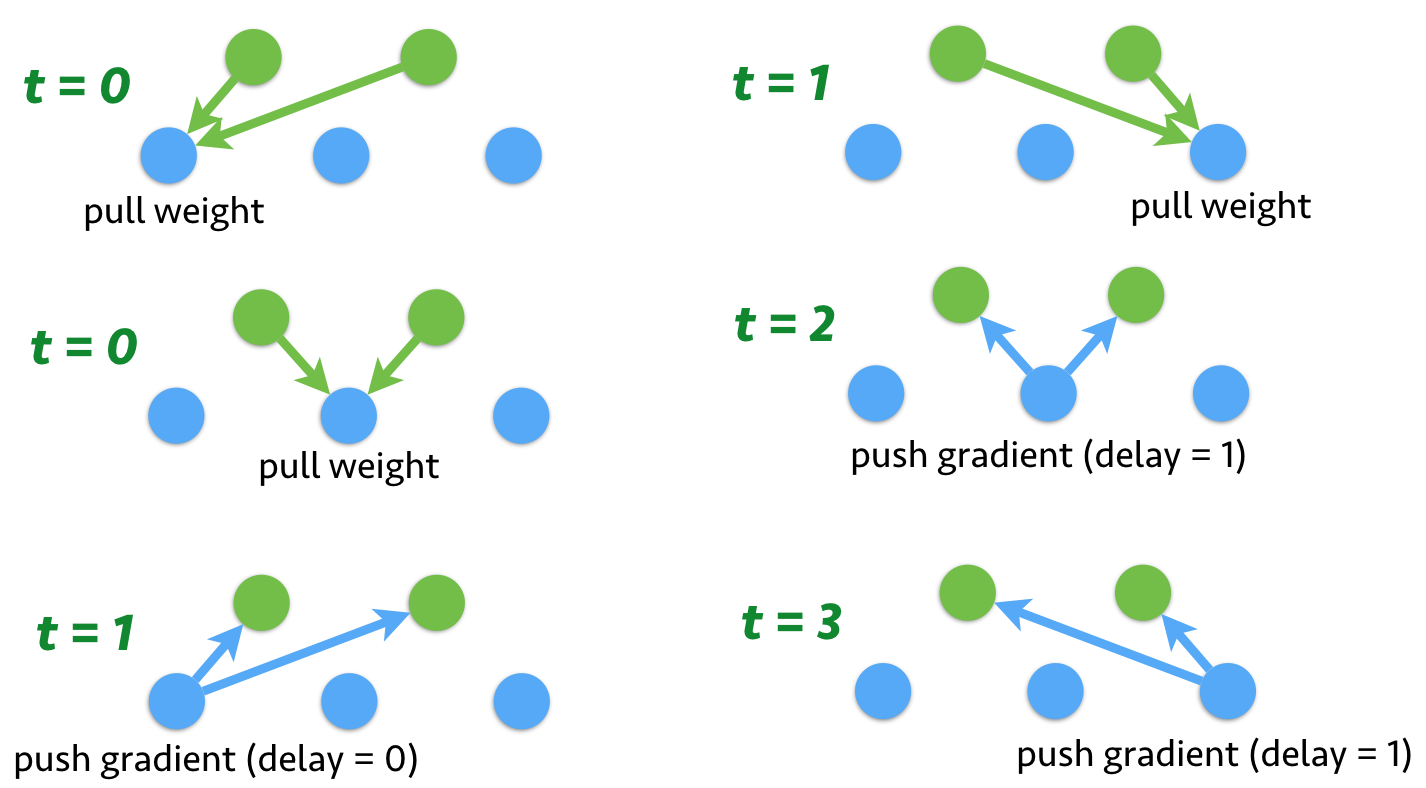}
 \caption{Asynchronous updates, where green nodes represent servers while blue correspond to workers. Adapted from \cite{dmlc2016pslite}\label{fig:async-sgd}}
\end{figure*}

Asynchronous algorithms such as Hogwild!\cite{recht2011hogwild} overcome the above problems. In asynchronous updates, fast workers do not wait for slow workers. One worker may be sending its local gradients to the server while others are calculating their gradients, as shown in Fig. \ref{fig:async-sgd}. For example, the first worker calculates and pushes its local gradients when $t=1$ whereas the second worker pushes gradients when $t=2$, although they both pull parameters from servers at the same timestamp ($t=0$).

The primary challenge raised by asynchronous updates is data staleness, because fast workers may always use stale parameters which jeopardizes model's convergence. In addition, the fastest worker is equivalent to update its local parameters by its sub-dataset, which causes the local model to deviate from the global model. To overcome the drawbacks of asynchronous updates, researchers have attempted to limit the staleness of the parameters. In bounded stale updates, fast workers will use stale parameters but the staleness (same as delay in Fig.\ref{fig:async-sgd}) is limited\cite{cipar2013solving,ho2013more}. The limitation on staleness mitigates the straggler problem to some extent and increases training throughput. However, determining ways to choose the limitation on staleness is a question that is worth discussing because an excessively large value means completely asynchronous updates, while a small value is similar to synchronous updates.

\section{Algorithm Optimization}
\label{sec:algorithm}

In this section, we demonstrate how to reduce the communications overhead in distributed DNN training from an algorithm perspective. Algorithm optimizations include reducing communication rounds and volumes as well as increasing the computation-communication overlap ratio. These optimizations are independent of the underlying network, and most algorithms can run on top of various network infrastructures and protocols.

\subsection{Communication Rounds}

When using SGD to train deep neural networks, the entire training process usually consists of multiple epochs and iterations. In regard to parallelization, nodes often exchange data at the end of every iteration. One intuitive way to cut down the communication overhead is reducing the number of data exchanges, or the communication rounds. The number of rounds for a node is related to the batch size and the communication period. Larger batch sizes and more extended communication periods can both reduce the number of data exchanging rounds.

\subsubsection{Large Batch Training}

Batch size is a significant hyperparameter that controls the amount of data read by a node in each iteration. A large batch size usually better approximates the distribution of the input data and introduces fewer variances into the gradient estimates than a small batch. Additionally, a large batch of data will take a longer time to process and incur fewer model parameter update; this finding is primarily due to the relationship between batch size, the number of iterations, and training data size. The following equation shows that using a large batch size leads to a reduction in the number of iterations; hence, the parameter updates are infrequent.

\begin{equation}
 \label{eq:batch-size}
 \text{Batch Size}\times \text{\# of iterations} = \text{\# of training data size}
\end{equation}

\begin{figure*}[ht]
 \centering
 \includegraphics[width=0.8\textwidth]{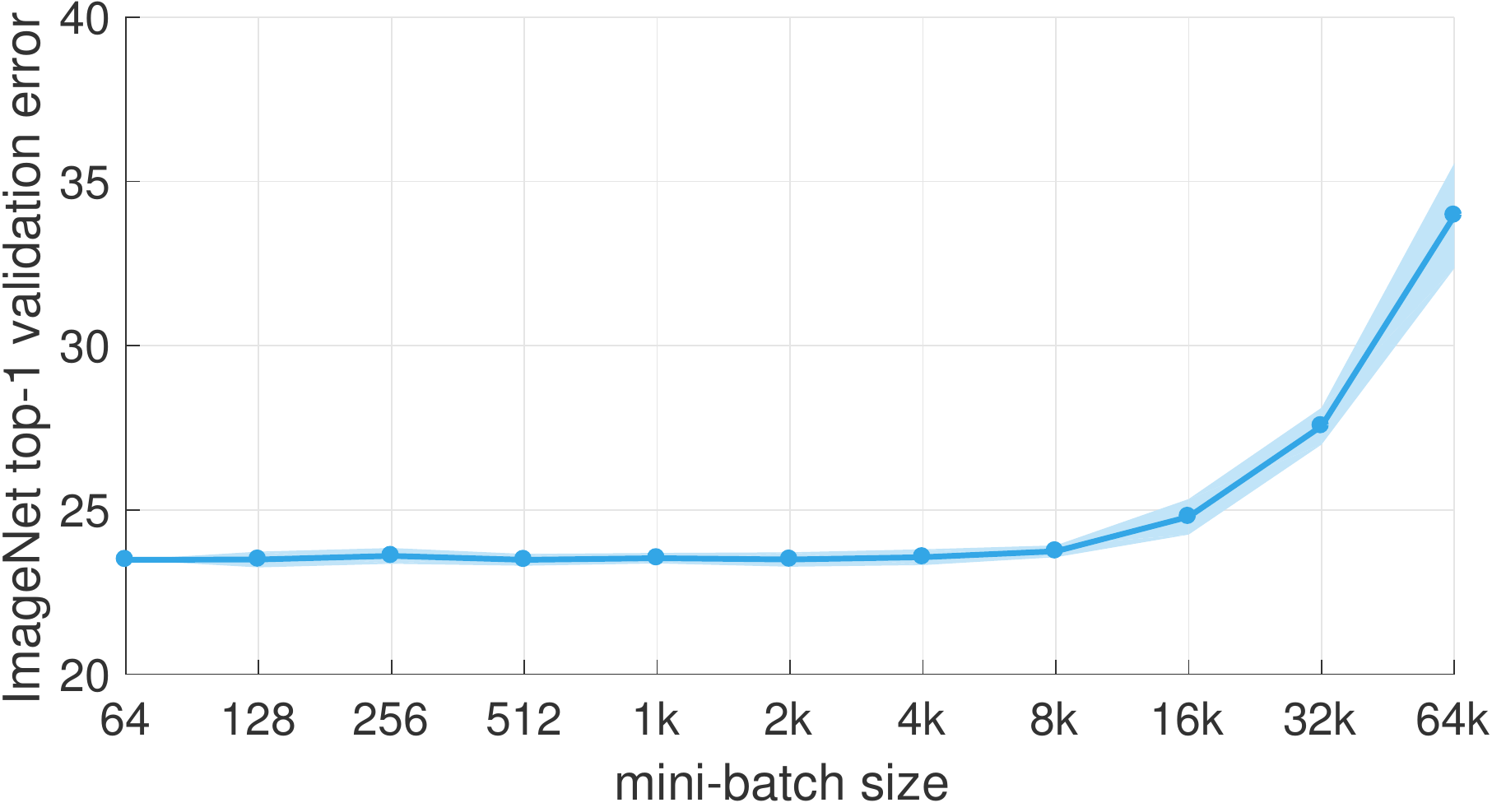}
 \caption{Example of large batch training of ResNet-50 on ImageNet. Adapted from \cite{goyal2017accurate}.}
 \label{fig:large-batch}
\end{figure*}

In distributed training circumstances with a data parallelism scheme, the batch size is the sum of all the local batch sizes of each node. Recall that in conventional distributed deep learning, nodes exchange gradients and model parameters at the end of each iteration. Since the shape and size of gradients and model parameters only depend on the DNN itself, the single iteration communicated message size remains constant, as changing the batch size does not change their shape and size. Therefore, increasing batch sizes reduces the number of iterations and the communications rounds. Take ResNet-50 as an example, if we fix the number of epochs at 100 and set the batch size to 1024 on two machines, the entire training process requires $250,000$ iterations, in contrast, with a batch size of 8192 on 16 machines, only 15625 iterations are required\cite{you2019fast}. Table \ref{tab:resnet-comparison} shows ResNet-50 training results with different configurations.

\begin{table*}[!t]
 \centering 
 \setlength{\tabcolsep}{3pt}
 \caption{Compare ResNet-50 Training Detail with Different Wroks}
 \label{tab:resnet-comparison}
 \resizebox{\textwidth}{!}{
 \begin{tabular}{ccccc}
 \hline
  Works & Batch Size & Hardware & Top-1 Accuracy & Training Time \\
 \hline
 He et al.\cite{he2016deep} & 256 & Tesla P100 $\times$ 8 & 75.3\% (baseline) & 29h \\
 Goyal et al.\cite{goyal2017accurate} & 8k & Tesla P100 $\times$ 256 & 76.3\% & 1h \\
 Cho et al.\cite{cho2017powerai} & 8k & Tesla P100 $\times$ 256 & 75.0\% & 50min \\
 Smith et al.\cite{smith2017don} & 8k$\rightarrow$16k & TPU (256 tensorcores) & 76.1\% & 45min \\
 Codreamu et al.\cite{codreanu2017scale} & 32k & KNL $\times$ 1024 & 75.3\% & 42min \\
 You et al.\cite{you2017scaling} & 32k & KNL $\times$ 2048 & 75.4\% & 20min \\
 Akiba et al.\cite{akiba2017extremely} & 32k & Tesla P100 $\times$ 1024 & 74.9\% & 15min \\
 Jia et al.\cite{jia2018highly} & 64k & Tesla P40 $\times$ 1024& 76.2\% & 8.7min \\
 Ying et al.\cite{ying2018image} & 32k& TPU $\times$ 1024 & 76.3\% &2.2min\\
 Mikami et al.\cite{mikami2018imagenet} & 54k & Tesla V100 $\times$ 3456 & 75.29\% & 2.0mins\\
 Yamazaki et al.\cite{yamazaki2019yet} & 80k & Tesla V100 $\times$ 2048 & 75.08\% & 1.2mins \\
 \hline
 \end{tabular}}
\end{table*}

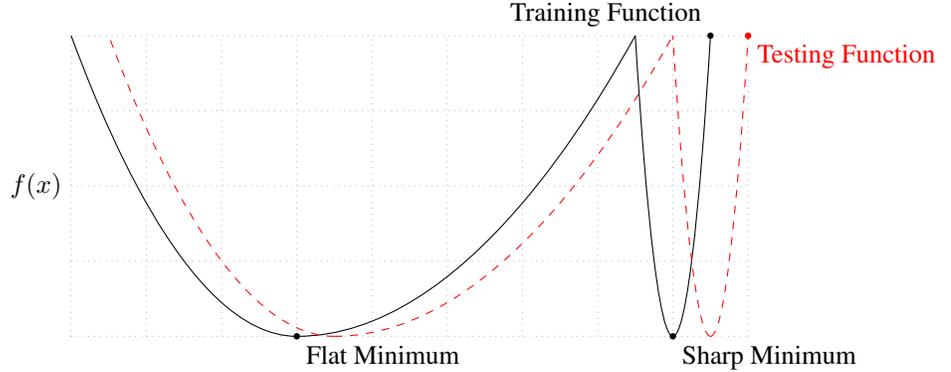
\begin{figure*}[htp]
 \flushright
 \begin{tikzpicture}
 \draw[step=1cm,gray,very thin,dotted] (-3,0) grid (6,4);
 \draw (0,0) parabola (4.5,4);
 \draw (0,0) parabola (-3,4);
 \draw (5,0) parabola (4.5,4);
 \draw (5,0) parabola (5.5,4);
 \draw[red,dashed] (0.5,0) parabola (5.0,4);
 \draw[red,dashed] (0.5,0) parabola (-2.5,4);
 \draw[red,dashed] (5.5,0) parabola (5.,4);
 \draw[red,dashed] (5.5,0) parabola (6.0,4);
 \filldraw[black] (0,0) circle (1pt) node[anchor=north west] {Flat Minimum};
 \filldraw[black] (5,0) circle (1pt) node[anchor=north west] {Sharp Minimum};
 \filldraw[black] (5.5,4) circle(1pt) node[anchor=south east] {Training Function};
 \filldraw[red] (6,4) circle(1pt) node[anchor=north west] {Testing Function};
 \filldraw[black] (-3,2) circle(0pt) node[anchor=east] {$f(x)$};
 \end{tikzpicture}
 \caption{A Conceptual Sketch of Flat and Sharp Minima. The Y-axis indicates value of the loss function and the X-axis the variables (parameters). Adapted from \cite{keskar2016large} \label{fig:cartoon}}
\end{figure*}

Nevertheless, by directly deploying a parallel SGD with a huge batch size, in practice it will likely suffer generalization ability degradations compared with training a small batch\cite{keskar2016large,goyal2017accurate}. Fig\ref{fig:large-batch}. illustrates this situation in which, when the batch size exceeds $8k$ (we use $1k$ to denote $1024$ samples), the error on the validation set increases dramatically as the batch size increases. The reason of this finding is that large batch methods tend to converge to a sharp minima of the training function\cite{keskar2016large}. These minima are characterized by a significant number of large positive eigenvalues in the Hessian matrix, indicating that the curvature around these minima is large. As shown in Fig.\ref{fig:cartoon}, sharp minima often generalize less well. Conversely, small batch method converge to flat minima, which is characterized by having many small eigenvalues for the Hessian matrix and a small curvature. Observations show that the loss function landscape of a DNN is such that large-batch methods are attracted to regions with sharp minima and are unable to escape from them\cite{keskar2016large}.

Many methods have been proposed to prevent models from converging into sharp minima, including learning rate scaling rules\cite{krizhevsky2014one, goyal2017accurate}, various warm-up schemes\cite{goyal2017accurate, you2019large}, and layerwise adaptive rate scaling (LARS)\cite{you2017scaling,you2018imagenet, you2019fast}. The scaling rules dictate that the learning rate should increase to prevent a loss of accuracy with an increase of the batch size. There are two rules of increasing learning rate: \emph{Sqrt Scaling Rule}\cite{krizhevsky2014one} and \emph{Linear Scaling Rule}\cite{goyal2017accurate}. When the batch size is multiplied by $k$, the first rule multiplies the learning rate by $\sqrt{k}$ to keep the variance of the gradient estimator constant\cite{krizhevsky2014one}, while the second rule multiplies the learning rate by $k$ based on an assumption introduced by \cite{goyal2017accurate}. When using the linear scaling rule, the learning rate usually is too large at the beginning for the model to converge. Hence, at the very beginning of training (i.e., 5 epochs), we slowly and smoothly increase the learning rate to the suggested value of the linear scaling rule. This scheme is called the gradual warm-up\cite{goyal2017accurate}. To deploy a warm-up on the RNN (i.e., LSTM and GRU) training process, You et al.\cite{you2019large} proposed a linear-epoch gradual warm-up (LEGW) scheme, in which the warm-up epochs are multiplied by $k$ when increasing the batch size $k$ times. LARS applies different learning rates for different layers in DNN because the distribution of gradients and parameters varies for different layers\cite{you2017scaling, you2018imagenet, you2019fast}. However, LARS performs poorly for attention models, indicating that its performance gains are not consistent across all tasks\cite{you2020large}. Therefore, You et al.\cite{you2020large} proposed a general layerwise adaptive large batch optimization technique called LAMB, which performs well across various tasks such as BERT\cite{devlin2018bert} and ResNet-50 training with minimal hyperparameter tuning. For RNN training, Chen et al.\cite{chen2016scalable} proposed a blockwise model update filtering (BMUF) to train LSTM on 16 GPUs and achieved a near-linear speedup.

\subsubsection{Periodic Communication}

Recall that when training DNNs using a traditional distributed SGD, communication often occurs at the end of each iteration. Suppose that we parallel a DNN training task on $K$ nodes with $T$ iterations, the complexity of rounds of a vanilla parallel SGD is $O (T)$. Due to such high complexity of communication rounds, many research studies suggest reducing the frequency of exchanging gradients and parameters.

\begin{figure*}[ht]
 \centering 
 \subfigure[Parallel SGD]{
  \includegraphics[width=0.8\textwidth, trim=0 210 0 210, clip]{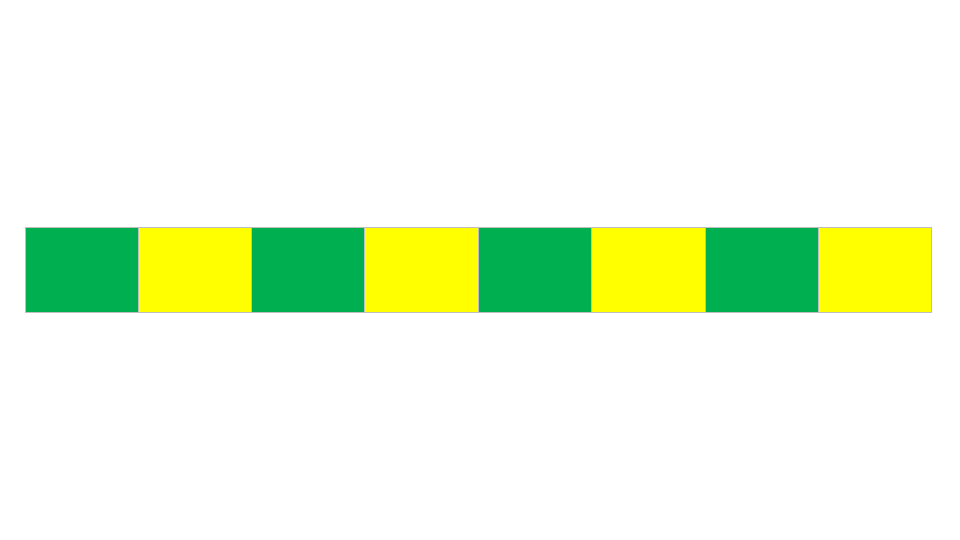}
  \label{fig:parallel-sgd}
 }
  
 \subfigure[Local SGD]{
  \includegraphics[width=0.8\textwidth, trim=0 210 0 210, clip]{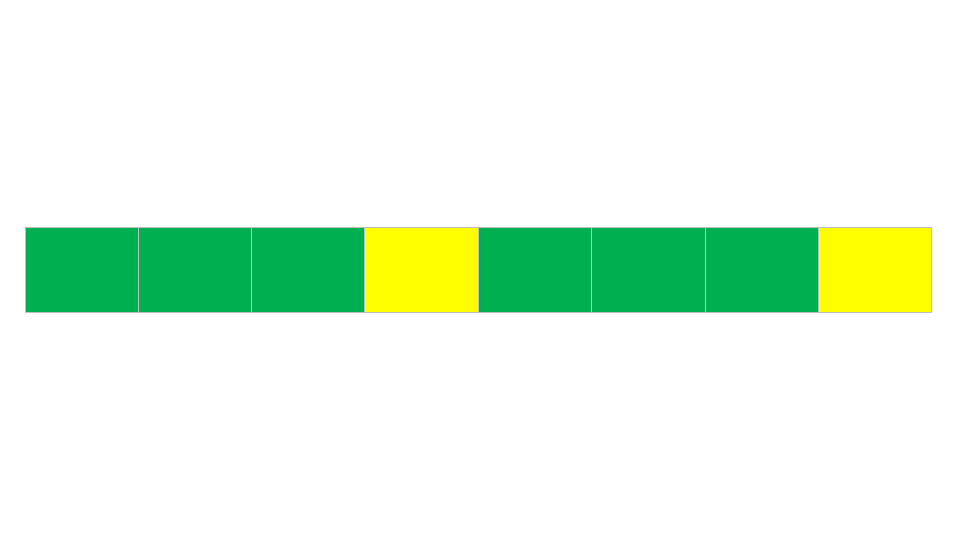}
 }
   
 \subfigure[One-Shot]{
  \includegraphics[width=0.8\textwidth, trim=0 210 0 210, clip]{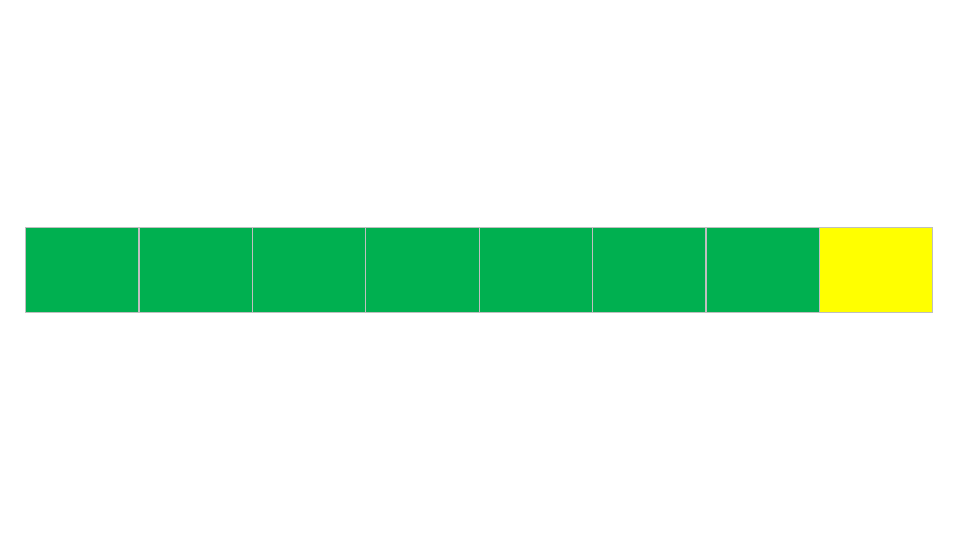}
 }
 \caption{Comparison of vanilla parallel SGD, local SGD and one-shot learning, where a green block indicates computation and a yellow block corresponds to communication.\label{fig:model-averaging}}
\end{figure*}

In model averaging, individual model parameters trained on local nodes are averaged periodically. We consider averaging operations occur at most for $\tau$ iterations, where $\tau$ is a factor of the period. Averaging occurs in every iteration, i.e., $\tau = 1$, which is identical to a vanilla parallel SGD. Conversely, averaging occurs only at the end of the training, i.e., $\tau = T$, which is equal to one-shot averaging. The case of $1 < \tau < T$ is another common setting. Fig. \ref{fig:model-averaging} depicts three cases. Experimental works\cite{zinkevich2010parallelized,mcdonald2010distributed,zhang2016parallel,su2015experiments} have verified that model averaging can reduce the communications overhead of the training time as long as the period is suitable. In addition, some theoretical studies\cite{yu2019parallel,stich2019local,arjevani2015communication,zhou2018convergence} have analyzed why model averaging can achieve good convergence rates.

\begin{table}[t!]
 \caption{Comparison of Periodic Averaging SGD\label{tab:periodic}}
 \resizebox{\textwidth}{!}{
 \begin{tabular}{ccc}
  \hline
 Works & Communication Rounds ($T/\tau$) & Linear Speedup\\
\hline
vanilla Parallel SGD\cite{dekel2012optimal} & $O (T)$ &\checkmark\\
One-Shot\cite{zinkevich2010parallelized,mcdonald2010distributed} & $O (1)$ &\\
K-AVG\cite{zhou2018convergence} & $O (T/\tau)$ &\\
PR-SGD\cite{yu2019parallel} & $O (K^{3/4}T^{3/4})$ &\checkmark when $T > N^3$\\
Local SGD\cite{stich2019local}& $O (K^{1/2}T^{1/2})$ & \checkmark\\
RI-SGD\cite{haddadpour2019trading}& $O (K^{1/2}T^{1/2})$ & \\
LUPA-SGD\cite{haddadpour2019local}& $O (K^{1/3}T^{1/3})$ & \checkmark\\
COCOD-SGD\cite{shen2019faster}& $O (K^{3/4}T^{3/4})$ & \checkmark\\
\hline
 \end{tabular}}
\end{table}

The one-shot averaging, which only requires communication at the end of the training, is an extreme case of model averaging. It has been shown that one-shot averaging has satisfying performance on both convex\cite{zinkevich2010parallelized} and some nonconvex\cite{mcdonald2010distributed} optimization problems. However, Zhang et al.\cite{zhang2016parallel} reported that certain nonconvex optimization problems cannot be solved by one-shot averaging. As a remedy, these researchers suggested that more frequent averaging can improve performance. 

Between one-shot averaging and vanilla parallel SGD, tremendous works have focused on periodical averaging, in which communication occurs every $\tau$ iterations. There is another aspect of research which attempts to reduce communication cost by adding redundancy into a training dataset. For instance, Haddadpour et al.\cite{haddadpour2018cross} show that by adding a proper amount of redundancy through coding theoretic means, a linear regression can be solved through one round of communication. Additionally, Haddadpour et al.\cite{haddadpour2019trading} demonstrate that by properly infusing redundancy into the training data with model averaging, it is conceivable to significantly reduce the order of communication rounds as PR-SGD\cite{yu2019parallel} with less accuracy degradation for a general nonconvex optimization. The advantages of model averaging have been examined from a practical point of view in\cite{su2015experiments}. Specifically, these findings show that model averaging performs well empirically in terms of reducing communication costs for a given accuracy. Arjevani et al.\cite{arjevani2015communication} studied the theoretical lower bounds of the number of communication rounds required for different settings in order to solve distributed convex learning and optimization problems.

In addition to model averaging, an asymmetrical \emph{push/pull} operation can also increase the communication period. Dean et al.\cite{dean2012large} proposed a feasible solution called asymmetrical push/pull, in which workers request update parameters from servers every $n_{fetch}$ step, and send gradients to servers every $n_{push}$ step ($n_{fetch}$ may not equal to $n_{push}$). However, they only provide experimental results for asymmetrical \emph{push/pull}, and the convergence analysis is not provided. Chen et al.\cite{chen2018lag} developed Lazily Aggregated Gradient (LAG) methods, in which workers and servers synchronize gradients and parameters only when some conditions are violated. These conditions are generated and exported by the estimation of the objective value descent for one iteration. LAG has convergence guarantees both experimentally and theoretically. When using nine workers to train a linear regression model, LAG can decrease the parameter server communication complexity from 5283 to 1756 , compared with the vanilla parallel SGD.

\subsection{Gradients Compression}

Each computation node requires communication operations to exchange gradients and model parameters during the distributed training. Using 32 bits variables to represent each element in the gradients and model parameters is the most commonly used configuration. When the size of the gradients and model parameters is large, the communication bottleneck caused by exchanging a large amount of 32 bits single-precision variables impairs the advantages of parallel SGDs. For example, consider training the BERT architecture for neural language processing tasks with approximately 340 million parameters, which implies that each full precision exchange between nodes is over 1.2GB\cite{devlin2018bert}. Hence, this finds is why we discuss ways to reduce the size of the message in this section.

\begin{figure*}[ht]
 
  \centering
  \input{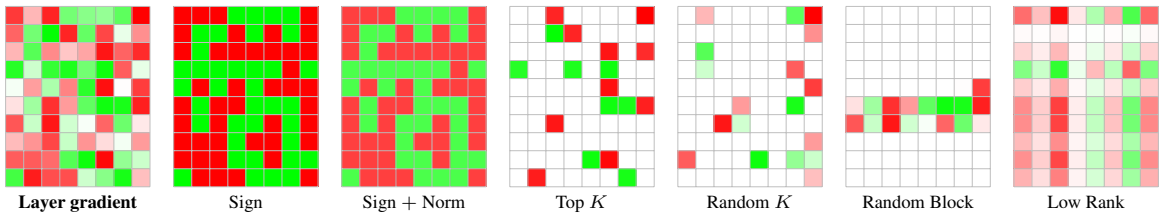}
  \caption{
   Different compression schemes, the first is one layer's original gradient as a matrix, while others are the output of various compression schemes taking the first as input. Red means negative values and green means positive. Adapted from \cite{vogels2019powersgd}}.
  \label{fig:compression-methods}
\end{figure*}

Gradient compression reduces the communication traffic by compressing the gradients transmitted during training. It is easy to implement and performs very well on some DNN models. However, the model's performance will be affected by these lossy compression methods, which is unacceptable in some areas (i.e., recommendation systems) where the accuracy is sensitive. Currently there are mainly three bandwidth-efficient compression approaches: quantization, sparsification and decomposition. The first, substitutes 32 bits variables for low-precision variables (i.e., 8 bits, 4 bits, or even 1 bit), whereas the second only transports important variables to avoid unnecessary overhead. Since these two approaches are orthogonal, they can be combined to compress the communication further. The third is not as popular as the first two; it transmits small matrices instead of large matrices by a matrix decomposition. Fig. \ref{fig:compression-methods} compares various compression methods.

\subsubsection{Quantization}
As a low-precision data representation method, quantization first discretizes continuous values and maps them to different integers in a range. A standard method of quantization is to use the sign of each element to represent the gradient. As far as we know, one-bit SGD\cite{seide20141} has pioneered the research of communication quantization for distributed deep learning. Recently, a similar algorithm called signSGD\cite{bernstein2018signsgd} was proposed for non-convex optimization.

Because the difference between gradients processed by sign-based quantization and original gradients is too large, the quantized gradient is often a biased estimate of the original gradient\cite{karimireddy2019error}, which makes the model converge slowly with a significant accuracy loss\cite{grubic2018synchronous}. To address this issue, we use the error feedback technique to correct the deviations of direction accumulated in the previous iterations. Error feedback maintains a vector $\textbf{\textit{e}}$ to store the accumulated difference between the quantized and the original gradients. The following equations give the rationale of the error feedback, where subscript $t$ represents the $t$ th iteration, $\beta$ is a decaying factor\cite{wu2018error} and $\text{Q}(\cdot)$ is the quantizer:

\begin{subequations}
 \begin{align}
  \textbf{\textit{e}}_{t+1}&= \textbf{\textit{g}}_t - \tilde{\textbf{\textit{g}}}_{t}\\
 \tilde{\textbf{\textit{g}}}_{t+1} &= \text{Q}(\textbf{\textit{g}}_{t+1} + \textbf{\textit{e}}_{t+1})
 \end{align}
\end{subequations}

Using error feedback, one-bit SGD achieves $10\times$ speed in training a speech DNN with little accuracy loss\cite{seide20141}. Karimireddy et al.\cite{karimireddy2019error} fixed signSGD using error feedback techniques and performed a theoretical analysis of the convergence of the method on nonconvex smooth functions.

Although error feedback can prevent the model from significant accuracy loss, gradients processed by such aggressive sign-based quantizer are still biased\cite{stich2018sparsified}. The other way to cope with biased gradients is to avoid producing them; hence, many unbiased methods have been proposed. These methods introduce randomness into the quantization operator, meaning that the quantized value of each element is subject to one probability distribution. A function $Q_s:\mathbb{R}^d\rightarrow\mathbb{R}^d$ that maps a deterministic input to a random vector with a  quantization level $s$ is called a stochastic quantizer, if the following holds for $\forall v \in \mathbb{R}^d$: (1)unbiased quantization$\mathbb{E}[Q_s (v)] = v$; (2)bounded variance$\mathbb{E}[\|Q_s (v)\|^2] \le (1+\beta_{d,s}) \|v\|^2$ where $\beta_{d,s}$ is a function of $d$ and $s$. Due to the extra randomness introduced by stochastic quantization, such quantizers must be an unbiased estimate of the original gradients and have a variance bound that ensures convergence as in a vanilla SGD.

In two parallel works, TernGrad\cite{wen2017terngrad} and QSGD\cite{alistarh2017qsgd}, both use stochastic unbiased gradients. For a given gradient $\textbf{\textit{g}}\in\mathbb{R}^d$, TernGrad\cite{wen2017terngrad} compresses each values into three levels $\{-1,0,+1\}$ by the following function:

\begin{equation}
 \tilde{\textbf{\textit{g}}} = \text{ternarize}(\textbf{\textit{g}}) = s \cdot \text{sign}(\textbf{\textit{g}}) \circ \textbf{\textit{b}}
\end{equation}

where
\begin{equation}
 s \triangleq\max(\text{abs}(\textbf{\textit{g}}))\triangleq \|\textbf{\textit{g}}\|_{\infty}
\end{equation}

is the maximum norm of vector $\textbf{\textit{g}}$, and $\textbf{\textit{b}}$ is a random binary vector that follows a Bernoulli distribution\cite{wen2017terngrad}. It can be proven that the quantized gradient $\tilde{\textbf{\textit{g}}}$ using a ternary function is an unbiased estimate of original gradient\cite{wen2017terngrad}. QSGD\cite{alistarh2017qsgd} has a similar but more sophisticated quantization function and provides a hyperparameter $s \ge 1$ to control the representation's precision levels. A smaller $s$ means less communication overhead with a higher level of information loss, whereas a larger $s$ means a lower level of information loss but more communication overhead.

Bidirectional quantization is often used in parameter server architecture to reduce \emph{worker-server} communication in both directions. SignSGD\cite{bernstein2018signsgd} applies the majority vote on the server-side to enable double quantization. TernGrad\cite{wen2017terngrad} offloads the model update operation to workers, and only the compressed gradient is transmitted between the workers and the server. In contrast, Asy-LPG\cite{yu2019double} quantizes the gradients and the model parameters in \emph{worker-server} and \emph{server-worker} direction respectively. 

To further reduce communication traffic, several works\cite{alistarh2017qsgd, wu2018error, lim20183lc} apply efficient lossless coding techniques (i.e., Elias coding\cite{elias1975universal}) after quantization. 

\subsubsection{Sparsification}

A limitation of quantization is that it can only compress a communication message up to 32 times, because we need at least one bit to represent numbers. However, sparsification does not have such a limitation; it only transmits values that play essential roles in the model update. A zero in the gradient means that the associated parameter is rarely updated, meaning that such gradient values have little effect on parameter updates. Therefore, it is a waste of bandwidth to transmit gradients that contains many zeros.

As far as we know, Strom et al.\cite{strom2015scalable} were the first to clip gradients via a static threshold, the element will not be transmitted if its absolute value below this static threshold. However, it is not very easy to select a reasonable threshold for various DNNs. To remedy this issue, subsequent works\cite{dryden2016communication, chen2018adacomp} applied local selections and dynamic threshold instead of a static threshold. A variant of threshold sparsification is the \emph{Top-K} method, in which each node only transmits $k$ largest (absolute) values\cite{aji2017sparse, lin2017deep, sattler2019sparse, renggli2018sparcml}. Theoretical work\cite{alistarh2018convergence} gave a convergence analysis of \emph{Top-K} sparsification under some assumptions, and pointed out that sparsification is equivalent to a stale update.

To ensure model convergence, we usually select values from gradient residuals, rather than manipulating the original gradients directly. The gradient residual is the sum of all previous gradients accumulated locally at each node\cite{strom2015scalable}. Stich et al.\cite{stich2018sparsified} proved that with accumulated gradients, sparsified SGD had the same convergence rates as vanilla parallel SGD. Deep gradient compression (DGC)\cite{lin2017deep} introduced momentum correction, local gradient clipping, momentum factor masking, and warm-up training to achieve better performance. By these methods, DGC compresses the gradient size of ResNet-50 from 97 MB to 0.35 MB without accuracy loss\cite{lin2017deep}. Wangni et al.\cite{wangni2018gradient} proposed random sparsification, which drops out indices of gradients randomly. To guarantee the sparsified vector is unbiased, the remaining part is appropriately amplified. Experiments show that random dropping causes little accuracy losses of 3-layer CNNs on CIFAR-10 datasets.

Since quantization and sparsification are two orthogonal methods, they can be integrated for deep compression. The integration is straightforward based on a centralized architecture (Parameter Server). However, there is a challenge that needs to be resolved for sparsification in a decentralized setting. Recall that sparsification only transmits "important" values in the gradient, which means that each node may receive different nonzero indices (dimensions) for the same gradient\cite{ben2019demystifying}. Fortunately, there are a series of implementation\cite{zhao2014kylix,nguyen2019topology,fang2018redsync,renggli2018sparcml} for sparse communication. Using such sparse collective communication protocols, we can implement a sparse gradient exchange in a decentralized architecture and combine it with quantization.

\subsubsection{Matrix Decomposition}

In the field of compression, an emerging method is to decompose a large gradient matrix into several small matrices before transmission, and then reconstruct after receiving it. Transmitting two small matrices has less communication overhead than transmitting one huge matrix. This method is feasible because of the correlation between gradients. GradiVeQ\cite{yu2018gradiveq} exploited such linear correlations between CNN gradients and applied principal component analysis (PCA)\cite{wold1987principal} to reduce the gradients' dimensions. In addition, the proposed method GradiVeQ also enables direct aggregation of compressed gradients. ATOMO\cite{wang2018atomo} is a general compression method built on top of a singular value decomposition (SVD)\cite{golub1971singular}. For a given gradient, ATOMO will produce a random unbiased gradient with minimal variance\cite{wang2018atomo}. PowerSGD\cite{vogels2019powersgd} performs a low rank decomposition with error feedback, and avoids the computationally expensive SVD step in ATOMO\cite{wang2018atomo} to achieve better scalability.

\subsection{Computation-Communication Overlap}

Since the gradient is generated in order from the last layer to the first layer during back propagation, there is no need to wait for the calculation of the previous layer to complete before sending the gradient of the later layer. In other words, former layers’ computation is independent of the latter layers’ communication, and the latter layers’ parameters update is independent of the former layers\cite{zhang2017poseidon}. Hence, we can transmit the gradient of the $L_{th}$ layer while computing the $(L-1)_{th}$ layer's gradient. Computation-communication overlap can hide part of communication overhead. In fact, this approach only makes the communication time appear to be reduced and does not truly optimize the communication bottleneck. Therefore, this approach can combined easily with other optimization strategies.


Poseidon\cite{zhang2017poseidon} provided a wait-free backward propagation (WFBP) scheduling algorithm. WFBP makes each layer start its communication once its gradients are calculated after backward propagation. However, different layers may have different computation and communication times, which means Poseidon may not outperform than the FIFO scheduling on some specific network models.

\begin{figure*}[ht]
 \centering
 \includegraphics[width=0.8\textwidth]{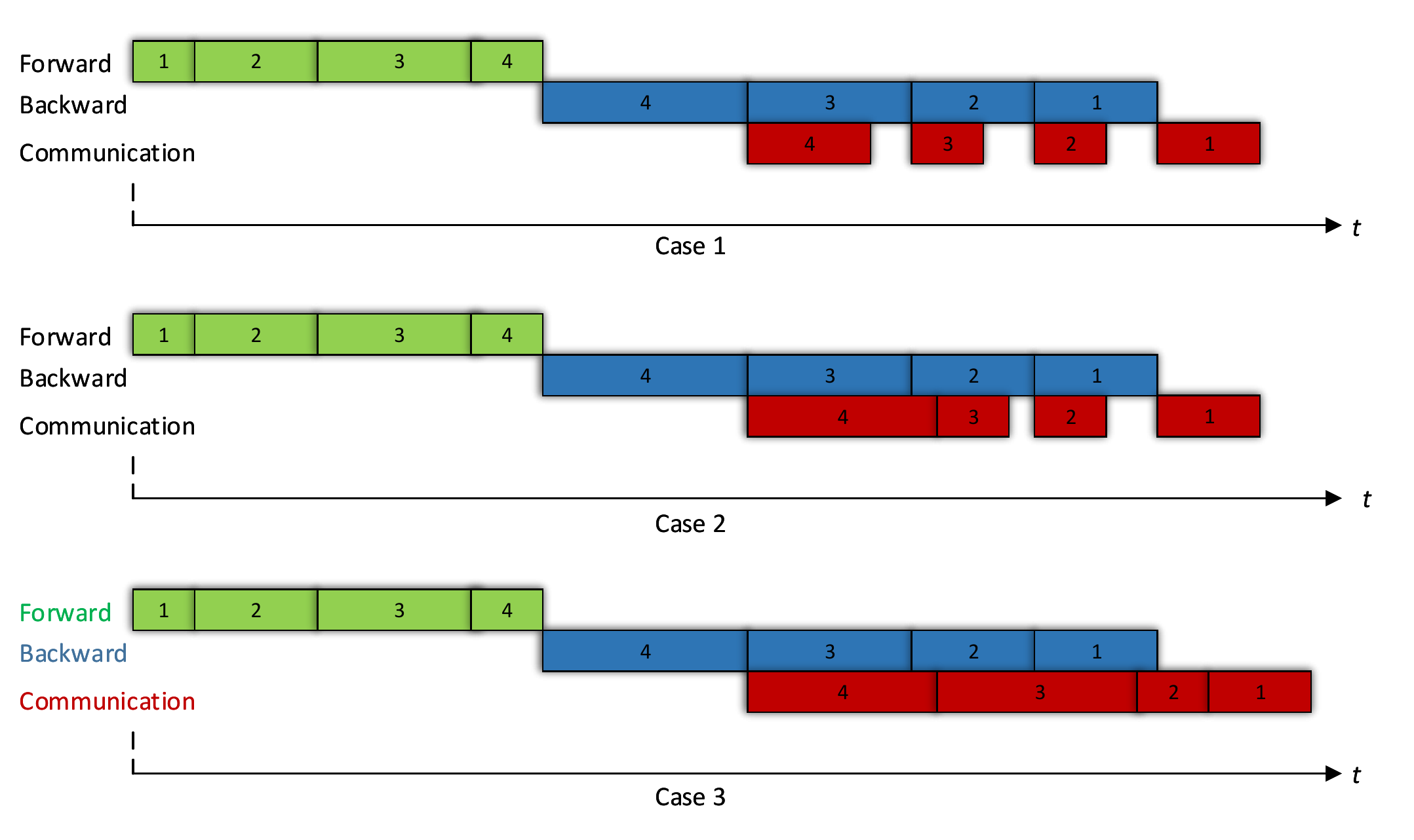}
 \caption{Three case of computation-communication overlap. Adapted from \cite{shi2019mg}.\label{fig:three-case}}
\end{figure*}

Shi et al.\cite{shi2019mg} pointed out that there are three cases of WFBP. Fig.\ref{fig:three-case} shows that both Case 1 and Case 2 are ideal cases in which we can easily hide the communication time. Case 3 could happened more often, especially in the high latency or low bandwidth network environment. They find that two or more small messages can be merged into one large message before being sent, and the merged gradients can be communicated with a smaller cost that is easily hidden by the computation. The newly proposed method, named merged-gradients WFBP (MG-WFBP), achieves much better scaling efficiency than WFBP.

Priority-based parameter propagation (P3)\cite{jayarajan2019priority} extended WFBP with priority-based scheduling, in which the first layer obtains the highest priority, and the last layer obtains the lowest priority. The tensor with the highest priority is processed first during the communication phase, regardless of when it is generated. P3 uses the tensor partition technique to handle Case 3, shown in Fig. \ref{fig:three-case}. Tensor partition technique splits layers’ parameter matrices into small proper pieces and assigns priorities to every slice based on their parent layers’ processing order in the forward propagation. Hashemi et al.\cite{hashemi2018tictac} exploited the execution order of computational graphs and proposed two heuristic scheduling algorithms: Timing-Independent Communication (TIC) and Timing-Aware Communication (TAC). Both algorithms are built on the properties of communication operations, such as communication dependency, communication time, and directly dependent compute load, etc. ByteScheduler\cite{peng2019generic} applies tensor partition and priority-based scheduling like P3, and designs a credit-based preemption approach to fully utilize the network's bandwidth. Credit-based preemption works similar to a sliding window, where credit is the window size. Small tensor pieces in the sliding window are sent simultaneously. ByteScheduler uses Bayesian optimization to find the ideal credit and partition sizes.

A recent work, one-step delay SGD (OD-SGD) \cite{xu2020od}, breaks the dependency between two consecutive iterations during training process, and improves distributed training performance by computation--communication overlap. The algorithm is more like a combination of synchronous and asynchronous SGD, which finds a balance between training speed and accuracy.

\section{Network Level Optimization}
\label{sec:network}
In this section, we mainly concentrate on optimizing low-level network infrastructures, including advanced centralized and decentralized architectures, messaging libraries, and network protocols. The benefits of network level optimization are intuitive. Modifying the low-level communication protocol will not have too much impact on the high-level training algorithm. However, network level optimization maybe not easy to implement, and the performance depends on the design of the distributed training system.

\subsection{Logical Architectures}
Centralized and decentralized architectures have different communication patterns and performance. We will discuss modern and advanced architectures in this subsection.
\subsubsection{Parameter Server}

\begin{figure*}[ht]
 \centering
 \subfigure[DistBelief]{
  \label{fig:ps-distbelief}
  \includegraphics[width=0.4\textwidth]{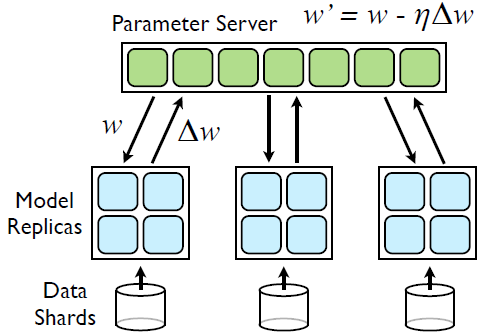}
 }
 \subfigure[Rudra]{
  \label{fig:ps-rudra}
  \includegraphics[width=0.4\textwidth]{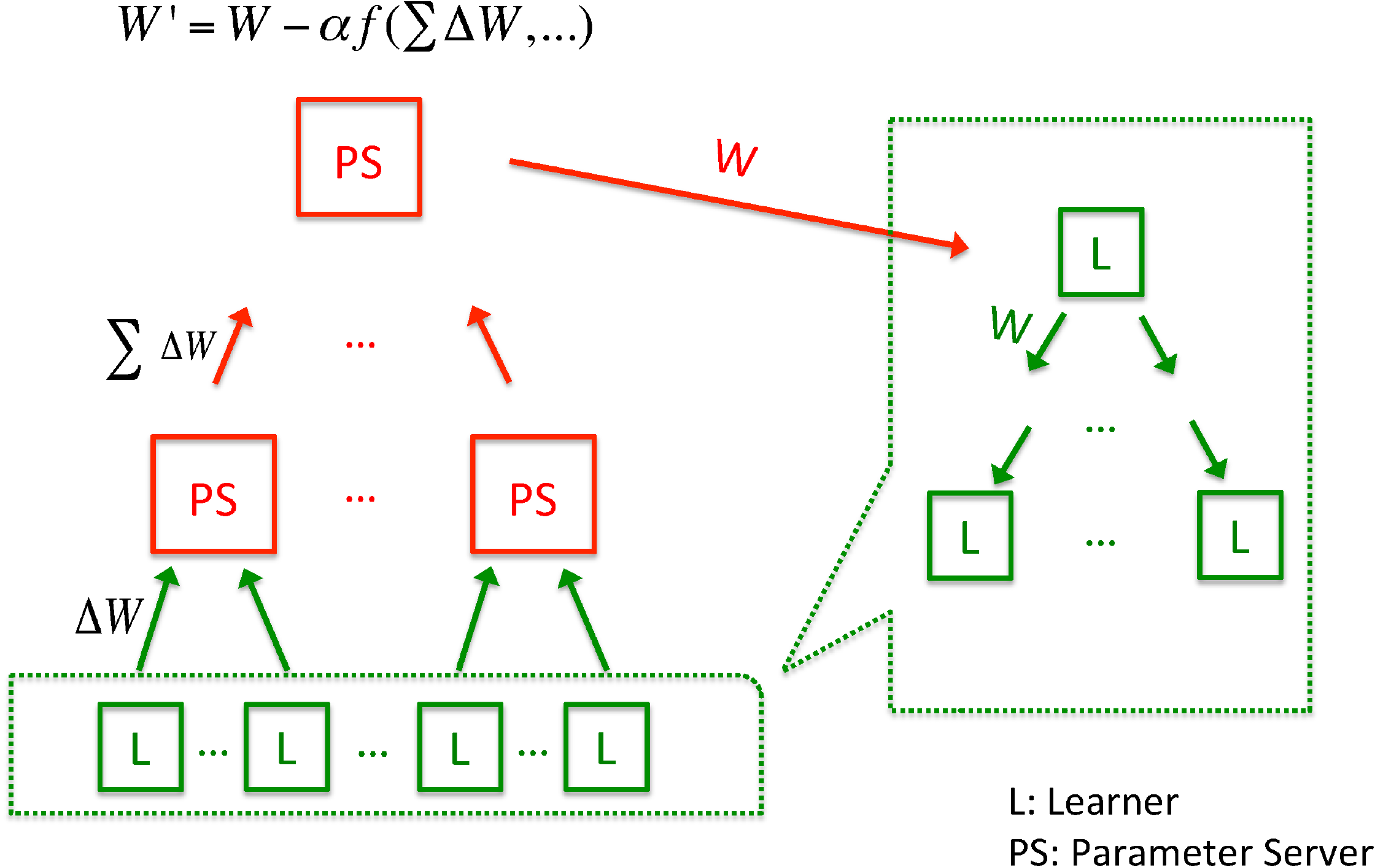}
 }
 \caption{Parameter server architectures.\label{fig:ps}}
\end{figure*}

Fig. \ref{fig:ps-distbelief} depicts the traditional parameter server architecture, in which the server is responsible for storing and updating global model parameters. The server is prone to network bottlenecks, especially when there are many working nodes. Tree-based parameter servers\cite{mai2015optimizing,gupta2016model,hsieh2017gaia} alleviate such bottlenecks to a certain extent. 

Mai et al.\cite{mai2015optimizing} treat each server as the root and build a spanning tree connecting all workers. All workers are leaf nodes in the spanning tree, whereas other nodes in the tree are servers. Each worker pushes gradients to their parents. The parents aggregate all received gradients and push the results upstream towards the root where the last step of aggregation is performed. The global weights are multicasted from the top root server down to the leaf worker. Through such a spanning tree, the communication overhead, in both push and pull operations is reduced. Gupta et al.\cite{gupta2016model} proposed similar tree-based architectures like\cite{mai2015optimizing}. To further alleviate the network traffic, the root server broadcasts the global weights directly down a tree constructed within all workers, see in Fig. \ref{fig:ps-rudra}. Heish et al.\cite{hsieh2017gaia} considered the physical distance of the parameter server and employed an intelligent communication mechanism over wide area networks (WAN) to efficiently utilize the bandwidth. In addition, some works such as Project Adam\cite{chilimbi2014project} and Geeps\cite{cui2016geeps} improve the throughput of parameter servers by caching and by isolated communications.

\subsubsection{Allreduce}

The key issue with traditional communication strategies is that, as the number of GPUs increases, the communication costs increase linearly. A classical implementation of Allreduce is a combination of Reduce operation followed by Broadcast which sends the result from the root to all processes. This implies a bottleneck on the root process. The optimized algorithms are based on a few principles: recursive vector halving, recursive vector doubling, recursive distance halving, recursive distance doubling, binary blocks, and ring\cite{rabenseifner2004optimization}.

\begin{figure*}[ht]
 \centering
 \includegraphics[width=0.8\textwidth]{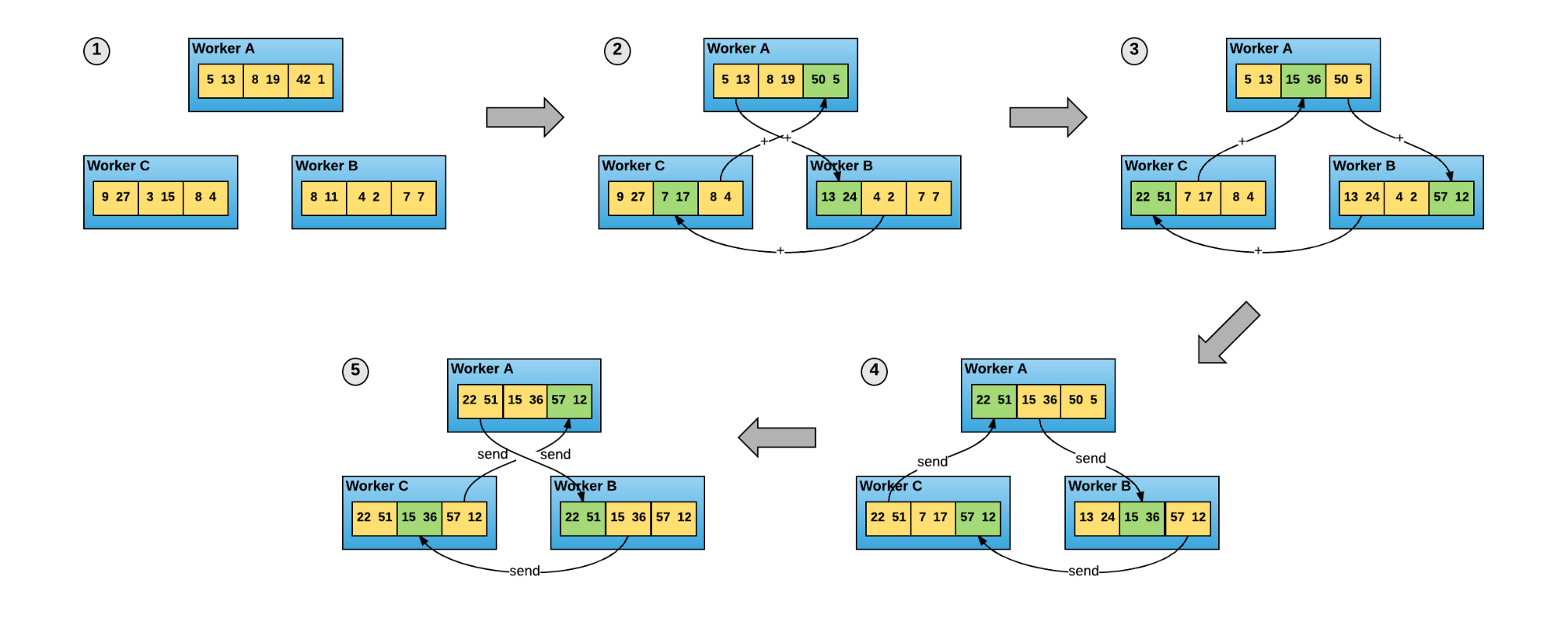}
 \caption{The Ring Allreduce algorithm. Adapted from \cite{sergeev2018horovod}.\label{fig:ringallreduce}}
\end{figure*}

To the best of our knowledge, Baidu first introduceda  ring-based Allreduce into distributed deep learning\cite{baidu2017allreduce}. A Ring Allreduce is made of two phases: Reduce-Scatter and Allgather; each phase includes $p-1$ communication steps when we use $p$ GPUs, see in Fig.\ref{fig:ringallreduce}. Each GPU maintains its local gradients, which are equally divided into $p$ chunks. In the reduce-scatter phase, each node sends and receives different chunks of a stored tensor. For the received chunk, each node adds it to the corresponding position in the buffer. After $p-1$ steps, each node holds a different part of the global result. In the all-gather phase, each node sends the part of the global result maintained by itself and receives other parts of the global result from other nodes. Each node holds a complete global result after $p-1$ steps. Hence, Ring Allreduce needs a total of $2 (p-1)$ communication steps. The complete communication process is described in Fig. As early as 2009, Patarasuk et al.\cite{patarasuk2009bandwidth} proved that a ring-based Allreduce has the optimal bandwidth of the Allreduce algorithms.

\begin{figure*}[ht]
 \centering
 \includegraphics[width=0.8\textwidth, trim=10 10 10 10, clip]{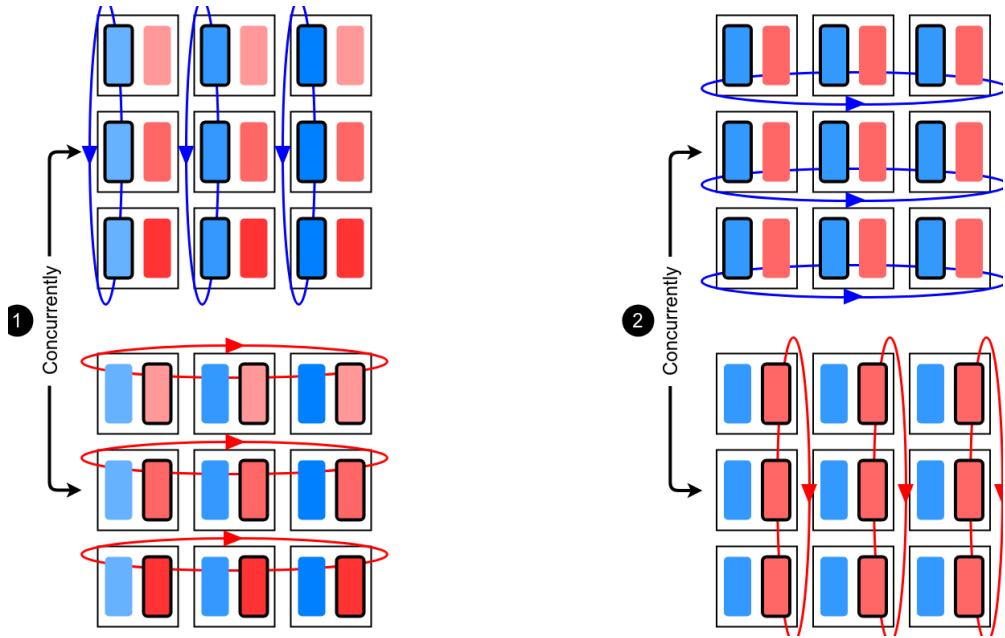}
 \caption{2D-Mesh Allreduce across a hypothetical $3 \times 3$ torus. Adapted from \cite{ying2018image}}
 \label{fig:2dtorus}
\end{figure*}

Mikami et al.\cite{mikami2018imagenet} proposed a 2D-Torus Allreduce topology in which the GPUs are arranged in a 2D grid. Each row contains $ph$ GPUs while each column contains $pv$ GPUs. There are three phases in  a 2D-Torus Allreduce: Reduce-Scatter, vertical Allreduce, and Allgather. Although a 2D-Torus Allreduce has one phase more than a  Ring Allreduce, its overall communication overhead is still smaller because $ph$ and $pv$ are less than $p$\cite{mikami2018imagenet}. A similar study by Ying et al.\cite{ying2018image} aggregates gradients in two phases with a 2D-Mesh topology, which utilizes two parallel ring-based reductions, each summing different halves of the payload along the horizontal and the vertical dimensions, see in Fig. \ref{fig:2dtorus}. The 2D-Mesh algorithm has twice the throughput of gradient aggregations than the 1D Ring Allreduce\cite{ying2018image}.

\begin{figure*}[ht]
 \centering
 \subfigure[Intra Ring]{
  \includegraphics[width=0.3\textwidth, trim=200 20 200 20, clip]{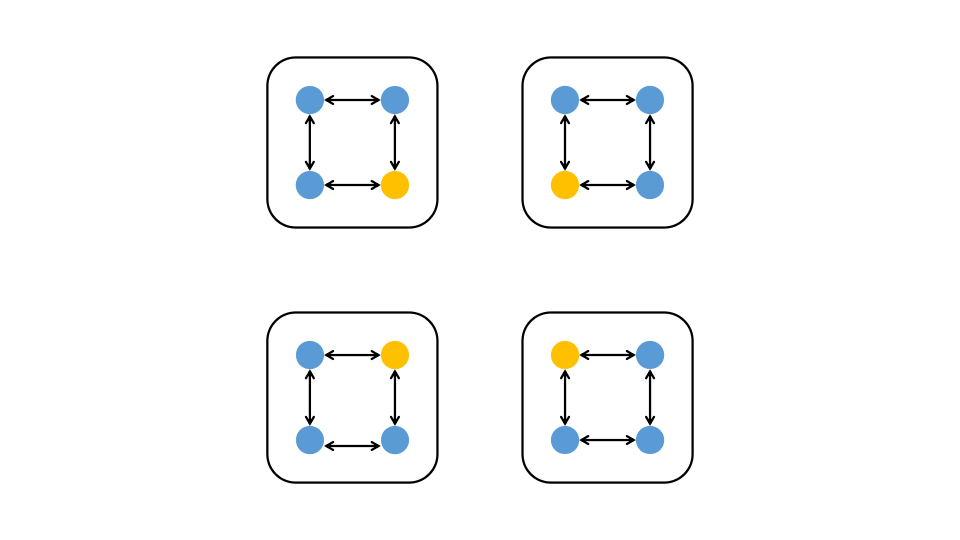}
 }
 \subfigure[Inter Ring]{
  \includegraphics[width=0.3\textwidth, trim=200 20 200 20, clip]{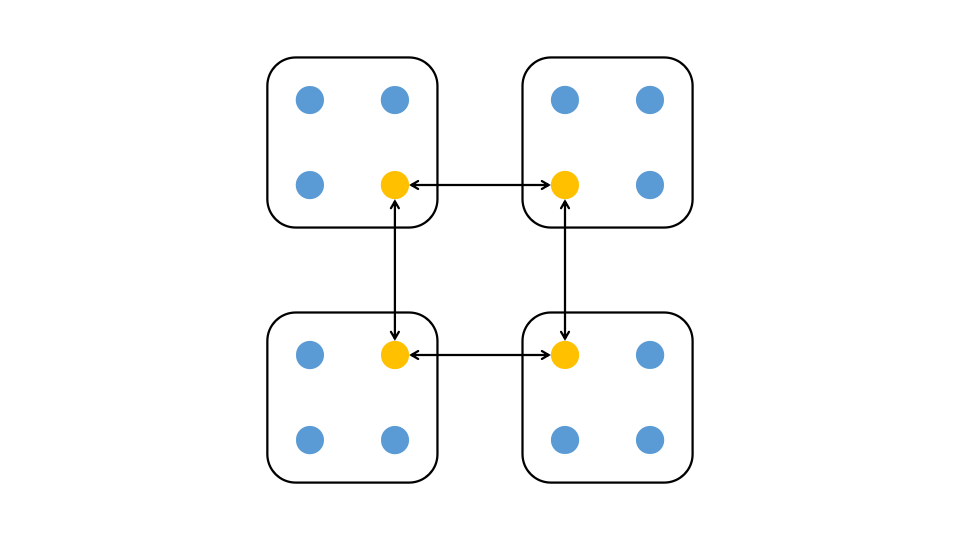}
 }
 \subfigure[Broadcast]{
  \includegraphics[width=0.3\textwidth, trim=200 20 200 20, clip]{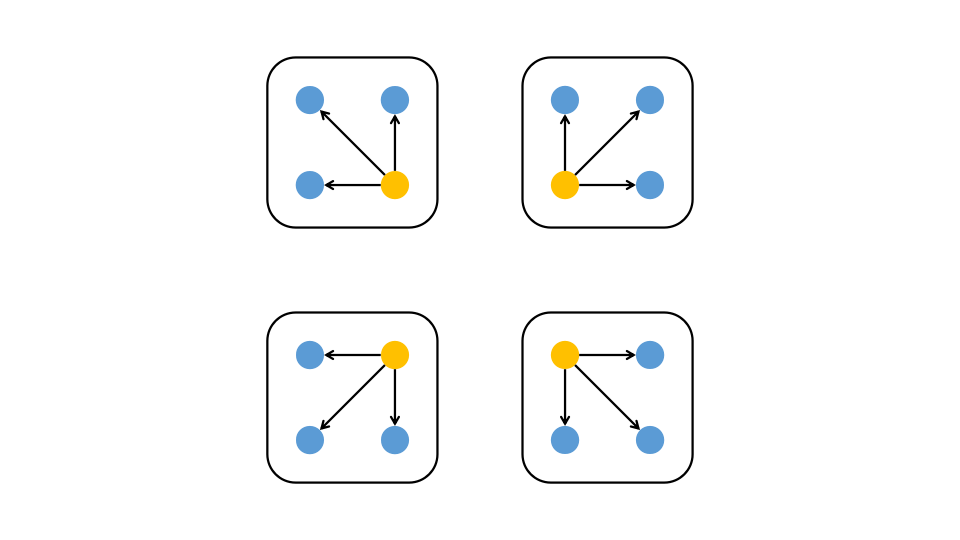}
 }
 \caption{Three phases of hierachical Allreduce, where each orange entry is local master node in each group.\cite{jia2018highly}\label{fig:hallreduce}}
\end{figure*}

Jia et al.\cite{jia2018highly} proposed hierarchical Allreduce to resolve the problem for a small tensor communication. These researchers split all $p$ GPUs into several groups and conducted a three-phase reduction. Fig. \ref{fig:hallreduce} shows the three-phase operations. The first phase is a separate ring Allreduce operation in independent groups, each of which consists of $k$ GPUs. In the second phase, a master node from each group operates Ring Allreduce to get a global result. Finally, the master node in each group broadcasts the global result to every GPU in its group. Compared with a Ring Allreduce, this three-phase hierarchical Allreduce decreases the running steps from $2 (p-1)$ to $4 (k-1)+2 (p/k-1)$.

\subsection{Message Level Libraries}
The performance of different parameter server systems and various Allreduce algorithms partially depends on the implementation of the message communication libraries. The parameter server is usually run on top of ZeroMQ\cite{hintjens2013zeromq} or gRPC\cite{google2019grpc}. ZeroMQ is a high performance and low latency asynchronous messaging library that supports multiple communication patterns, and gRPC is a high-performance remote procedure call (RPC) framework developed by Google. As far as we know, the most commonly used message communication libraries in the parameter server are still ZeroMQ and gRPC.

At present, there are many message level communication libraries that implement various Allreduce algorithms and other collective communication algorithms efficiently, including MPI\cite{gropp2002mpich2}, Gloo\cite{facebook2017gloo}, NCCL\cite{jeaugey2017nccl}, Baidu Allreduce\cite{baidu2017allreduce}, Aluminum\cite{dryden2018aluminum} and BlueConnect\cite{cho2019blueconnect}. Thanks to the high performance of MPI, there are many optimizations based on MPI\_Allreduce, including Horovod\cite{sergeev2018horovod}, MXNet-MPI\cite{mamidala2018mxnet} and TensorFlow-MPI\cite{vishnu2016distributed}. To reduce communications, Horovod uses tensor fusion that sends several small tensors simultaneously\cite{sergeev2018horovod}. NCCL implements multi-GPU and multinode collective communication primitives that are optimized for NVIDIA GPUs\cite{jeaugey2017nccl}. A series of studies by Awan et al.\cite{awan2018optimized, awan2016efficient} optimized Bcast operation based on NCCL and CUDA-Aware MPI, respectively. BlueConnect\cite{cho2019blueconnect} decomposes a single Allreduce operation into a series of parallelizable Reduce-Scatter and Allgather operations to reduce the communication cost. Experiments show that the performance of BlueConnect incurs less communication overhead than Gloo and Baidu Allreduce with more GPUs\cite{cho2019blueconnect}.

\subsection{Network Protocols}

\begin{figure*}[ht]
 \centering
 \subfigure[TCP/IP Socket]{
  \label{fig:tcpip}
  \includegraphics[width=0.3\textwidth, trim=350 50 350 50, clip]{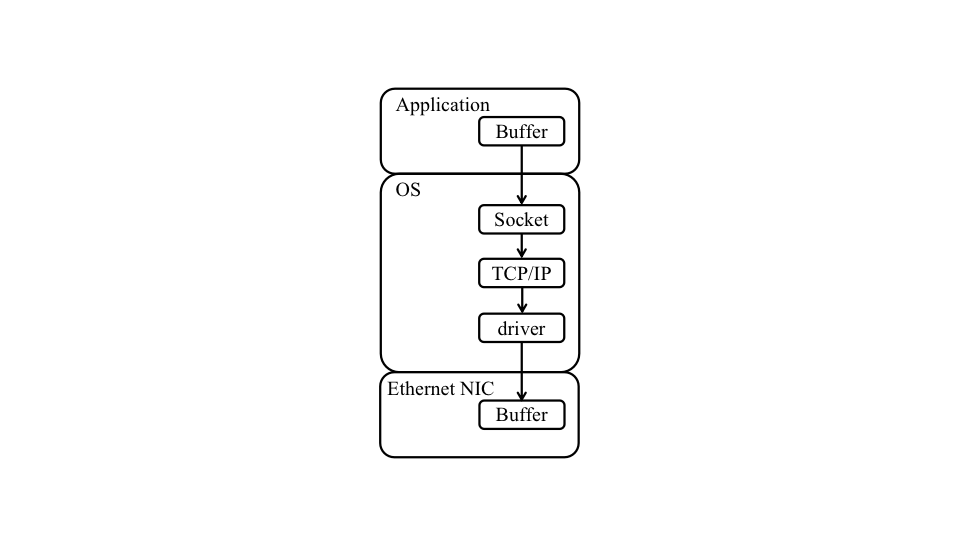}
 }
 \subfigure[IPoIB]{
  \label{fig:ipoib}
  \includegraphics[width=0.3\textwidth, trim=350 50 350 50, clip]{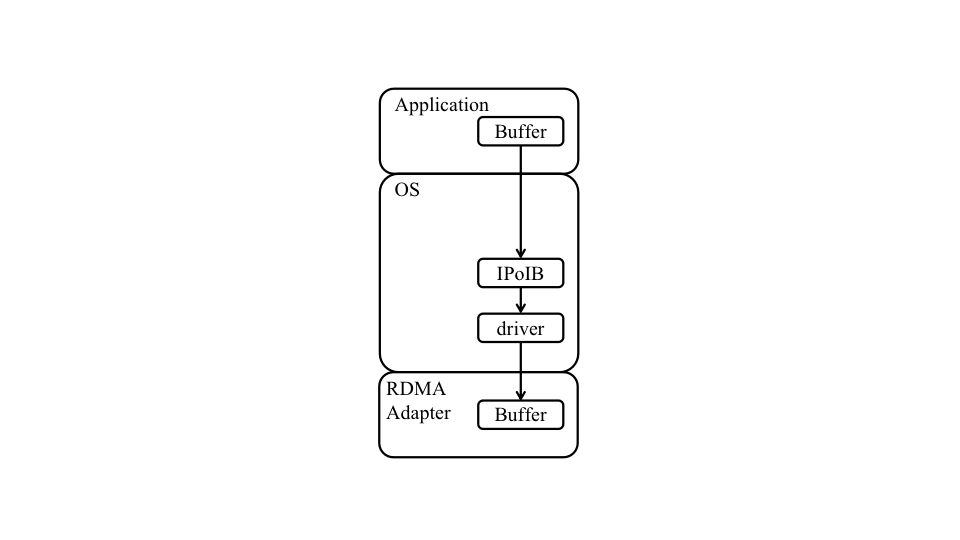}
 }
 \subfigure[RDMA]{
  \label{fig:rdma}
  \includegraphics[width=0.3\textwidth, trim=350 50 350 50, clip]{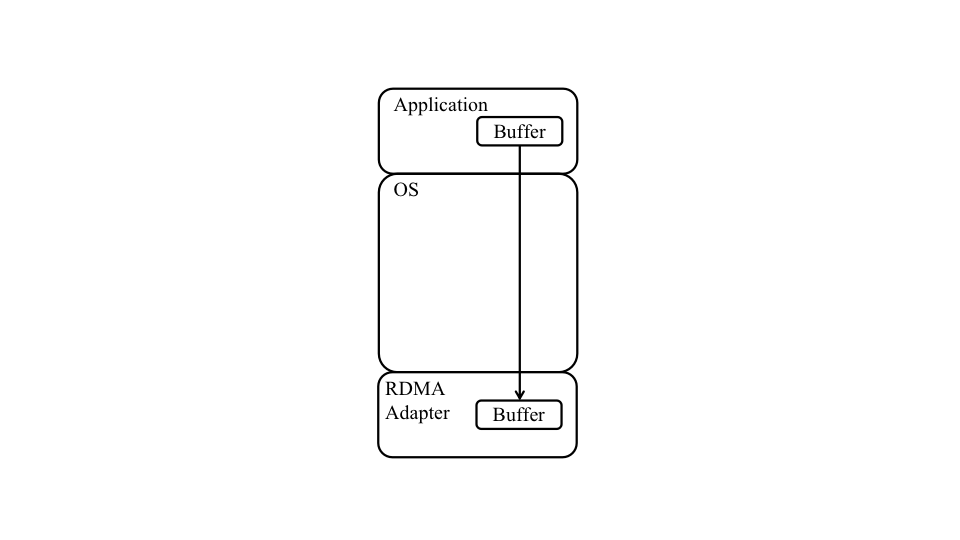}
 }
 \caption{Comparison of Socket, IPoIB and RDMA, where RDMA Adapter is typically InfiniBand.\label{fig:protocols}}
\end{figure*}

Traditional message level communication libraries are implemented based on the TCP/IP protocol, which handles data sending and receiving by sockets. Each node must create a socket object and establish a connection to the receiver before sending data. Data are processed by the operating system and encapsulated with different protocol headers until its copied into the network interface controller (NIC) buffer, as shown in Fig. \ref{fig:tcpip}. Such operations are wasteful in distributed training that needs low network latency. Therefore, high performance and low latency networks that run on top of associated network hardware (i.e., InfiniBand) attract much research attention; the two most common are remote direct memory access (RDMA)\cite{recio2007remote} and internet protocol over InfiniBand (IPoIB)\cite{chu2006transmission}. As illustrated in Fig.\ref{fig:rdma}, RDMA allows one machine to directly read and write the memory of another machine without involving the operating system, which enables high performance and low latency networking. The network interface in RDMA is called verbs, which provide two types of communication paradigms: \emph{message} and \emph{memory}. IPoIB, as the name implies, encapsulates IP datagrams over an InfiniBand card and enables TCP/IP applications to run on top of InfiniBand without any code modification\cite{chu2006transmission}. However, as shown in Fig.\ref{fig:ipoib}, IPoIB cannot bypass the host operating system like RDMA in Fig. \ref{fig:rdma}.

With the advent of RDMA and IPoIB, tremendous works have been done on improving the performance of distributed training systems, including MXNet\cite{li2019improving,liu2018optimizing}, TensorFlow\cite{xue2019fast,biswas2018accelerating,jia2018improving}, CNTK\cite{banerjee2016re,chu2017efficient}, Caffe\cite{awan2017s} and the IBM deep learning platform\cite{ren2017irdma}, to leverage its high bandwidth and low latency. The memory communication paradigm is the most popular method due to its low memory demands\cite{ren2017irdma,xue2019fast}. Some works\cite{xue2019fast,chu2017efficient,ren2017irdma} consider communications between GPUs and explore GPU direct RDMA (GDR), which allows an RDMA NIC to access GPU memory directly without going through host memory. Gradients are aggregated in GPUs by using the GDR technique. Furthermore, Biswas et al.\cite{biswas2018accelerating} designed an adaptive RDMA-based gRPC to dynamically adjust communication mechanisms for different message sizes in a deep learning workload.

According to experimental results\cite{xue2019fast,li2019improving,jia2018improving}, using RDMA and IPoIB to replace the TCP/IP protocol can significantly speed up training. In addition, RDMA performs better than IPoIB in distributed training\cite{li2019improving, liu2018optimizing}. Experiments by Liu et al.\cite{liu2018optimizing} report that compared with IPoIB (53\%), the RDMA-capable network achieves near-linear (96\%) speedup when scaling Inception-v3 training on 100 GPUs. The result also fits the previous description of IPoIB and RDMA verbs.

Recently, Xia et al.\cite{xia2019rethinking} rethought the distributed training process of DNNs and designed a bounded loss tolerance transmission protocol, which ignores a fraction of the packet loss during training. The new protocol achieves shorter job completion times than the traditional TCP/IP protocol. However, the random dropping rate has to be tuned when changing the DNN architecture. It is challenging to choose a reasonable dropping rate for various DNNs.

\section{Conclusion Remarks}
\label{sec:conclusion}
It has been shown that communications overhead is a significant obstacle to achieving a desirable performance for distributed DNN training. In this paper, we provide a comprehensive survey of the recent research on communication optimization techniques for distributed DNN training. Both theoretical and experimental studies are investigated. We divide these communication optimization strategies into two dimensions: algorithm level optimization and network level optimization. From the algorithm perspective, we elaborate on techniques to reduce communication volumes and computation-communication overlaps. In terms of network level optimization, we discussed the impact of different topologies and network protocols on distributed deep learning. Below, we highlight potential new research directions and challenges.

\emph{Focusing on different deep learning tasks and neural network models.} At present, there are a considerable number of communication optimizations that focus on image classification tasks, especially ResNet-50 trained on ImageNet dataset. Natural language processing and recommendation system related tasks have not received notable research attention. For example, the large embedding matrix in the recommendation system model may become a new bottleneck in communication optimization.

\emph{Local SGD for nonconvex problems.} For periodic communications, there still remains some research opportunities for nonconvex problems. Despite many theoretical works on model averaging, the research on whether the linear speedup with $\tau > 1$ can be preserved for nonconvex optimizations is still unexplored. Individually, the lower bounds on the number of communication rounds for nonconvex optimizations to achieve linear speedup is an interesting research direction. 

\emph{Trade-off between model accuracy and compression ratio.} The core challenge that requires consideration in gradient compression is the trade-off between model accuracy and the compression ratio. Conventional approaches to prevent models from diverging include error feedback (for quantization) and local gradient accumulation (for sparsification). More advanced methods such as squared error feedback\cite{lim20183lc} need to be explored in future works.

\emph{Higher computation-communication overlap ratio.} The ratio of computation to communication is essential for deploying pipeline training. For the sake of a higher overlap rating, various algorithms use different strategies to arrange communication operations and shrink communication time. However, these algorithms are generally heuristic and achieve nonoptimal solutions in such scheduling problems. Better optimization algorithms, such as dynamic programming, maybe be suitable for solving this problem.

\emph{Large-scale DNN training on top of different datacenter network topology.} The physical topology of the datacenter network has a significant impact on distributed deep learning. A recent work by Wang et al.\cite{wang2018bml} deployed a parameter server over BCube\cite{guo2009bcube} instead of a traditional Fat-Tree\cite{al2008scalable} architecture and achieved good performance on LeNet-5\cite{lecun1998gradient} and VGG-19\cite{simonyan2014very} training. Other network topologies, such as DCell\cite{guo2008dcell}, may further accelerate large-scale neural network training.

\emph{Communications overhead analysis tools are required.} Another critical research issue is the performance model and measurement tools of distributed DNN training. Performance models allow us to theoretically analyze the various costs of distributed training\cite{yan2015performance, qi2017paleo}, and the measurement tools help us study communication behavior and find the bottlenecks in the distributed training tasks. Although some 
deep learning frameworks including Tensorflow\cite{abadi2016tensorflow} and MXNet\cite{chen2015mxnet} currently provide performance analysis tools, these tools cannot analyze network behavior. Advanced tools for network analysis such as Horovod timeline\cite{sergeev2018horovod} and SketchDLC\cite{xu2019sketchdlc} are still required.


\bibliographystyle{unsrt}  

\bibliography{reference}

\begin{thebibliography}{100}

\bibitem{chen2015deepdriving}
Chenyi Chen, Ari Seff, Alain Kornhauser, and Jianxiong Xiao.
\newblock Deepdriving: Learning affordance for direct perception in autonomous
  driving.
\newblock In {\em Proceedings of the IEEE International Conference on Computer
  Vision}, pages 2722--2730, 2015.

\bibitem{chollet2017xception}
Fran{\c{c}}ois Chollet.
\newblock Xception: Deep learning with depthwise separable convolutions.
\newblock In {\em Proceedings of the IEEE conference on computer vision and
  pattern recognition}, pages 1251--1258, 2017.

\bibitem{zhang2020resnest}
Hang Zhang, Chongruo Wu, Zhongyue Zhang, Yi~Zhu, Zhi Zhang, Haibin Lin, Yue
  Sun, Tong He, Jonas Mueller, R~Manmatha, et~al.
\newblock Resnest: Split-attention networks.
\newblock {\em arXiv preprint arXiv:2004.08955}, 2020.

\bibitem{cho2014learning}
Kyunghyun Cho, Bart van Merrienboer, Caglar Gulcehre, Dzmitry Bahdanau, Fethi
  Bougares, Holger Schwenk, and Yoshua Bengio.
\newblock Learning phrase representations using rnn encoder--decoder for
  statistical machine translation.
\newblock In {\em Proceedings of the 2014 Conference on Empirical Methods in
  Natural Language Processing (EMNLP)}, pages 1724--1734, 2014.

\bibitem{brown2020language}
Tom~B. Brown, Benjamin Mann, Nick Ryder, Melanie Subbiah, Jared Kaplan,
  Prafulla Dhariwal, Arvind Neelakantan, Pranav Shyam, Girish Sastry, Amanda
  Askell, Sandhini Agarwal, Ariel Herbert-Voss, Gretchen Krueger, Tom Henighan,
  Rewon Child, Aditya Ramesh, Daniel~M. Ziegler, Jeffrey Wu, Clemens Winter,
  Christopher Hesse, Mark Chen, Eric Sigler, Mateusz Litwin, Scott Gray,
  Benjamin Chess, Jack Clark, Christopher Berner, Sam McCandlish, Alec Radford,
  Ilya Sutskever, and Dario Amodei.
\newblock Language models are few-shot learners.
\newblock 2020.

\bibitem{covington2016deep}
Paul Covington, Jay Adams, and Emre Sargin.
\newblock Deep neural networks for youtube recommendations.
\newblock In {\em Proceedings of the 10th ACM conference on recommender
  systems}, pages 191--198. ACM, 2016.

\bibitem{hochreiter1997long}
Sepp Hochreiter and J{\"u}rgen Schmidhuber.
\newblock Long short-term memory.
\newblock {\em Neural computation}, 9(8):1735--1780, 1997.

\bibitem{szegedy2015going}
Christian Szegedy, Wei Liu, Yangqing Jia, Pierre Sermanet, Scott Reed, Dragomir
  Anguelov, Dumitru Erhan, Vincent Vanhoucke, and Andrew Rabinovich.
\newblock Going deeper with convolutions.
\newblock In {\em Proceedings of the IEEE conference on computer vision and
  pattern recognition}, pages 1--9, 2015.

\bibitem{he2016deep}
Kaiming He, Xiangyu Zhang, Shaoqing Ren, and Jian Sun.
\newblock Deep residual learning for image recognition.
\newblock In {\em Proceedings of the IEEE conference on computer vision and
  pattern recognition}, pages 770--778, 2016.

\bibitem{russakovsky2015imagenet}
Olga Russakovsky, Jia Deng, Hao Su, Jonathan Krause, Sanjeev Satheesh, Sean Ma,
  Zhiheng Huang, Andrej Karpathy, Aditya Khosla, Michael Bernstein, et~al.
\newblock Imagenet large scale visual recognition challenge.
\newblock {\em International journal of computer vision}, 115(3):211--252,
  2015.

\bibitem{lindholm2008nvidia}
Erik Lindholm, John Nickolls, Stuart Oberman, and John Montrym.
\newblock Nvidia tesla: A unified graphics and computing architecture.
\newblock {\em IEEE micro}, 28(2):39--55, 2008.

\bibitem{jouppi2017datacenter}
Norman~P Jouppi, Cliff Young, Nishant Patil, David Patterson, Gaurav Agrawal,
  Raminder Bajwa, Sarah Bates, Suresh Bhatia, Nan Boden, Al~Borchers, et~al.
\newblock In-datacenter performance analysis of a tensor processing unit.
\newblock In {\em Proceedings of the 44th Annual International Symposium on
  Computer Architecture}, pages 1--12, 2017.

\bibitem{ben2019demystifying}
Tal Ben-Nun and Torsten Hoefler.
\newblock Demystifying parallel and distributed deep learning: An in-depth
  concurrency analysis.
\newblock {\em ACM Computing Surveys (CSUR)}, 52(4):65, 2019.

\bibitem{chahal2019hitchhiker}
Karanbir Chahal, Manraj~Singh Grover, Kuntal Dey, and Rajiv~Ratn Shah.
\newblock A hitchhiker’s guide on distributed training of deep neural
  networks.
\newblock {\em Journal of Parallel and Distributed Computing}, 2019.

\bibitem{mayer2020scalable}
Ruben Mayer and Hans-Arno Jacobsen.
\newblock Scalable deep learning on distributed infrastructures: Challenges,
  techniques, and tools.
\newblock {\em ACM Computing Surveys (CSUR)}, 53(1):1--37, 2020.

\bibitem{mittal2019survey}
Sparsh Mittal and Shraiysh Vaishay.
\newblock A survey of techniques for optimizing deep learning on gpus.
\newblock {\em Journal of Systems Architecture}, page 101635, 2019.

\bibitem{bottou2012stochastic}
L{\'e}on Bottou.
\newblock Stochastic gradient descent tricks.
\newblock In {\em Neural networks: Tricks of the trade}, pages 421--436.
  Springer, 2012.

\bibitem{bottou2010large}
L{\'e}on Bottou.
\newblock Large-scale machine learning with stochastic gradient descent.
\newblock In {\em Proceedings of COMPSTAT'2010}, pages 177--186. Springer,
  2010.

\bibitem{rumelhart1986learning}
David~E Rumelhart, Geoffrey~E Hinton, and Ronald~J Williams.
\newblock Learning representations by back-propagating errors.
\newblock {\em nature}, 323(6088):533--536, 1986.

\bibitem{gholami2018integrated}
Amir Gholami, Ariful Azad, Peter Jin, Kurt Keutzer, and Aydin Buluc.
\newblock Integrated model, batch, and domain parallelism in training neural
  networks.
\newblock In {\em Proceedings of the 30th on Symposium on Parallelism in
  Algorithms and Architectures}, pages 77--86, 2018.

\bibitem{dean2012large}
Jeffrey Dean, Greg Corrado, Rajat Monga, Kai Chen, Matthieu Devin, Mark Mao,
  Marc'aurelio Ranzato, Andrew Senior, Paul Tucker, Ke~Yang, et~al.
\newblock Large scale distributed deep networks.
\newblock In {\em Advances in neural information processing systems}, pages
  1223--1231, 2012.

\bibitem{li2014scaling}
Mu~Li, David~G Andersen, Jun~Woo Park, Alexander~J Smola, Amr Ahmed, Vanja
  Josifovski, James Long, Eugene~J Shekita, and Bor-Yiing Su.
\newblock Scaling distributed machine learning with the parameter server.
\newblock In {\em 11th $\{$USENIX$\}$ Symposium on Operating Systems Design and
  Implementation ($\{$OSDI$\}$ 14)}, pages 583--598, 2014.

\bibitem{li2014communication}
Mu~Li, David~G Andersen, Alexander~J Smola, and Kai Yu.
\newblock Communication efficient distributed machine learning with the
  parameter server.
\newblock In {\em Advances in Neural Information Processing Systems}, pages
  19--27, 2014.

\bibitem{lian2017can}
Xiangru Lian, Ce~Zhang, Huan Zhang, Cho-Jui Hsieh, Wei Zhang, and Ji~Liu.
\newblock Can decentralized algorithms outperform centralized algorithms? a
  case study for decentralized parallel stochastic gradient descent.
\newblock In {\em Advances in Neural Information Processing Systems}, pages
  5330--5340, 2017.

\bibitem{valiant1990bridging}
Leslie~G Valiant.
\newblock A bridging model for parallel computation.
\newblock {\em Communications of the ACM}, 33(8):103--111, 1990.

\bibitem{recht2011hogwild}
Benjamin Recht, Christopher Re, Stephen Wright, and Feng Niu.
\newblock Hogwild: A lock-free approach to parallelizing stochastic gradient
  descent.
\newblock In {\em Advances in neural information processing systems}, pages
  693--701, 2011.

\bibitem{dmlc2016pslite}
DMLC.
\newblock Ps-lite documents.
\newblock \url{https://ps-lite.readthedocs.io/}.

\bibitem{cipar2013solving}
James Cipar, Qirong Ho, Jin~Kyu Kim, Seunghak Lee, Gregory~R Ganger, Garth
  Gibson, Kimberly Keeton, and Eric Xing.
\newblock Solving the straggler problem with bounded staleness.
\newblock In {\em Presented as part of the 14th Workshop on Hot Topics in
  Operating Systems}, 2013.

\bibitem{ho2013more}
Qirong Ho, James Cipar, Henggang Cui, Seunghak Lee, Jin~Kyu Kim, Phillip~B
  Gibbons, Garth~A Gibson, Greg Ganger, and Eric~P Xing.
\newblock More effective distributed ml via a stale synchronous parallel
  parameter server.
\newblock In {\em Advances in neural information processing systems}, pages
  1223--1231, 2013.

\bibitem{goyal2017accurate}
Priya Goyal, Piotr Doll{\'a}r, Ross Girshick, Pieter Noordhuis, Lukasz
  Wesolowski, Aapo Kyrola, Andrew Tulloch, Yangqing Jia, and Kaiming He.
\newblock Accurate, large minibatch sgd: Training imagenet in 1 hour.
\newblock {\em arXiv preprint arXiv:1706.02677}, 2017.

\bibitem{you2019fast}
Yang You, Zhao Zhang, Cho-Jui Hsieh, Jim Demmel, and Kurt Keutzer.
\newblock Fast deep neural network training on distributed systems and cloud
  tpus.
\newblock {\em IEEE Transactions on Parallel and Distributed Systems}, 2019.

\bibitem{cho2017powerai}
Minsik Cho, Ulrich Finkler, Sameer Kumar, David Kung, Vaibhav Saxena, and
  Dheeraj Sreedhar.
\newblock Powerai ddl.
\newblock {\em arXiv preprint arXiv:1708.02188}, 2017.

\bibitem{smith2017don}
Samuel~L Smith, Pieter-Jan Kindermans, Chris Ying, and Quoc~V Le.
\newblock Don't decay the learning rate, increase the batch size.
\newblock {\em arXiv preprint arXiv:1711.00489}, 2017.

\bibitem{codreanu2017scale}
Valeriu Codreanu, Damian Podareanu, and Vikram Saletore.
\newblock Scale out for large minibatch sgd: Residual network training on
  imagenet-1k with improved accuracy and reduced time to train.
\newblock {\em arXiv preprint arXiv:1711.04291}, 2017.

\bibitem{you2017scaling}
Yang You, Igor Gitman, and Boris Ginsburg.
\newblock Scaling sgd batch size to 32k for imagenet training.
\newblock {\em arXiv preprint arXiv:1708.03888}, 6, 2017.

\bibitem{akiba2017extremely}
Takuya Akiba, Shuji Suzuki, and Keisuke Fukuda.
\newblock Extremely large minibatch sgd: training resnet-50 on imagenet in 15
  minutes.
\newblock {\em arXiv preprint arXiv:1711.04325}, 2017.

\bibitem{jia2018highly}
Xianyan Jia, Shutao Song, Wei He, Yangzihao Wang, Haidong Rong, Feihu Zhou,
  Liqiang Xie, Zhenyu Guo, Yuanzhou Yang, Liwei Yu, et~al.
\newblock Highly scalable deep learning training system with mixed-precision:
  Training imagenet in four minutes.
\newblock {\em arXiv preprint arXiv:1807.11205}, 2018.

\bibitem{ying2018image}
Chris Ying, Sameer Kumar, Dehao Chen, Tao Wang, and Youlong Cheng.
\newblock Image classification at supercomputer scale.
\newblock {\em arXiv preprint arXiv:1811.06992}, 2018.

\bibitem{mikami2018imagenet}
Hiroaki Mikami, Hisahiro Suganuma, et~al.
\newblock Imagenet/resnet-50 training in 224 seconds.
\newblock {\em arXiv preprint arXiv:1811.05233}, 2018.

\bibitem{yamazaki2019yet}
Masafumi Yamazaki, Akihiko Kasagi, Akihiro Tabuchi, Takumi Honda, Masahiro
  Miwa, Naoto Fukumoto, Tsuguchika Tabaru, Atsushi Ike, and Kohta Nakashima.
\newblock Yet another accelerated sgd: Resnet-50 training on imagenet in 74.7
  seconds.
\newblock {\em arXiv preprint arXiv:1903.12650}, 2019.

\bibitem{keskar2016large}
Nitish~Shirish Keskar, Dheevatsa Mudigere, Jorge Nocedal, Mikhail Smelyanskiy,
  and Ping Tak~Peter Tang.
\newblock On large-batch training for deep learning: Generalization gap and
  sharp minima.
\newblock {\em arXiv preprint arXiv:1609.04836}, 2016.

\bibitem{krizhevsky2014one}
Alex Krizhevsky.
\newblock One weird trick for parallelizing convolutional neural networks.
\newblock {\em arXiv preprint arXiv:1404.5997}, 2014.

\bibitem{you2019large}
Yang You, Jonathan Hseu, Chris Ying, James Demmel, Kurt Keutzer, and Cho-Jui
  Hsieh.
\newblock Large-batch training for lstm and beyond.
\newblock In {\em Proceedings of the International Conference for High
  Performance Computing, Networking, Storage and Analysis}, pages 1--16, 2019.

\bibitem{you2018imagenet}
Yang You, Zhao Zhang, Cho-Jui Hsieh, James Demmel, and Kurt Keutzer.
\newblock Imagenet training in minutes.
\newblock In {\em Proceedings of the 47th International Conference on Parallel
  Processing}, page~1. ACM, 2018.

\bibitem{you2020large}
Yang You, Jing Li, Sashank Reddi, Jonathan Hseu, Sanjiv Kumar, Srinadh
  Bhojanapalli, Xiaodan Song, James Demmel, Kurt Keutzer, and Cho-Jui Hsieh.
\newblock Large batch optimization for deep learning: Training bert in 76
  minutes.
\newblock In {\em International Conference on Learning Representations}, 2020.

\bibitem{devlin2018bert}
Jacob Devlin, Ming-Wei Chang, Kenton Lee, and Kristina Toutanova.
\newblock Bert: Pre-training of deep bidirectional transformers for language
  understanding.
\newblock {\em arXiv preprint arXiv:1810.04805}, 2018.

\bibitem{chen2016scalable}
Kai Chen and Qiang Huo.
\newblock Scalable training of deep learning machines by incremental block
  training with intra-block parallel optimization and blockwise model-update
  filtering.
\newblock In {\em 2016 ieee international conference on acoustics, speech and
  signal processing (icassp)}, pages 5880--5884. IEEE, 2016.

\bibitem{zinkevich2010parallelized}
Martin Zinkevich, Markus Weimer, Lihong Li, and Alex~J Smola.
\newblock Parallelized stochastic gradient descent.
\newblock In {\em Advances in neural information processing systems}, pages
  2595--2603, 2010.

\bibitem{mcdonald2010distributed}
Ryan McDonald, Keith Hall, and Gideon Mann.
\newblock Distributed training strategies for the structured perceptron.
\newblock In {\em Human Language Technologies: The 2010 Annual Conference of
  the North American Chapter of the Association for Computational Linguistics},
  pages 456--464. Association for Computational Linguistics, 2010.

\bibitem{zhang2016parallel}
Jian Zhang, Christopher De~Sa, Ioannis Mitliagkas, and Christopher R{\'e}.
\newblock Parallel sgd: When does averaging help?
\newblock {\em arXiv preprint arXiv:1606.07365}, 2016.

\bibitem{su2015experiments}
Hang Su and Haoyu Chen.
\newblock Experiments on parallel training of deep neural network using model
  averaging.
\newblock {\em arXiv preprint arXiv:1507.01239}, 2015.

\bibitem{yu2019parallel}
Hao Yu, Sen Yang, and Shenghuo Zhu.
\newblock Parallel restarted sgd with faster convergence and less
  communication: Demystifying why model averaging works for deep learning.
\newblock In {\em Proceedings of the AAAI Conference on Artificial
  Intelligence}, volume~33, pages 5693--5700, 2019.

\bibitem{stich2019local}
Sebastian~Urban Stich.
\newblock Local sgd converges fast and communicates little.
\newblock In {\em ICLR 2019 ICLR 2019 International Conference on Learning
  Representations}, number CONF, 2019.

\bibitem{arjevani2015communication}
Yossi Arjevani and Ohad Shamir.
\newblock Communication complexity of distributed convex learning and
  optimization.
\newblock In {\em Advances in neural information processing systems}, pages
  1756--1764, 2015.

\bibitem{zhou2018convergence}
Fan Zhou and Guojing Cong.
\newblock On the convergence properties of a k-step averaging stochastic
  gradient descent algorithm for nonconvex optimization.
\newblock In {\em Proceedings of the 27th International Joint Conference on
  Artificial Intelligence}, pages 3219--3227. AAAI Press, 2018.

\bibitem{dekel2012optimal}
Ofer Dekel, Ran Gilad-Bachrach, Ohad Shamir, and Lin Xiao.
\newblock Optimal distributed online prediction using mini-batches.
\newblock {\em Journal of Machine Learning Research}, 13(Jan):165--202, 2012.

\bibitem{haddadpour2019trading}
Farzin Haddadpour, Mohammad~Mahdi Kamani, Mehrdad Mahdavi, and Viveck Cadambe.
\newblock Trading redundancy for communication: Speeding up distributed sgd for
  non-convex optimization.
\newblock In {\em International Conference on Machine Learning}, pages
  2545--2554, 2019.

\bibitem{haddadpour2019local}
Farzin Haddadpour, Mohammad~Mahdi Kamani, Mehrdad Mahdavi, and Viveck Cadambe.
\newblock Local sgd with periodic averaging: Tighter analysis and adaptive
  synchronization.
\newblock In {\em Advances in Neural Information Processing Systems}, pages
  11080--11092, 2019.

\bibitem{shen2019faster}
Shuheng Shen, Linli Xu, Jingchang Liu, Xianfeng Liang, and Yifei Cheng.
\newblock Faster distributed deep net training: computation and communication
  decoupled stochastic gradient descent.
\newblock In {\em Proceedings of the 28th International Joint Conference on
  Artificial Intelligence}, pages 4582--4589. AAAI Press, 2019.

\bibitem{haddadpour2018cross}
Farzin Haddadpour, Yaoqing Yang, Viveck Cadambe, and Pulkit Grover.
\newblock Cross-iteration coded computing.
\newblock In {\em 2018 56th Annual Allerton Conference on Communication,
  Control, and Computing (Allerton)}, pages 196--203. IEEE, 2018.

\bibitem{chen2018lag}
Tianyi Chen, Georgios Giannakis, Tao Sun, and Wotao Yin.
\newblock Lag: Lazily aggregated gradient for communication-efficient
  distributed learning.
\newblock In {\em Advances in Neural Information Processing Systems}, pages
  5050--5060, 2018.

\bibitem{seide20141}
Frank Seide, Hao Fu, Jasha Droppo, Gang Li, and Dong Yu.
\newblock 1-bit stochastic gradient descent and its application to
  data-parallel distributed training of speech dnns.
\newblock In {\em Fifteenth Annual Conference of the International Speech
  Communication Association}, 2014.

\bibitem{bernstein2018signsgd}
Jeremy Bernstein, Yu-Xiang Wang, Kamyar Azizzadenesheli, and Animashree
  Anandkumar.
\newblock Signsgd: Compressed optimisation for non-convex problems.
\newblock In {\em International Conference on Machine Learning}, pages
  559--568, 2018.

\bibitem{karimireddy2019error}
Sai~Praneeth Karimireddy, Quentin Rebjock, Sebastian Stich, and Martin Jaggi.
\newblock Error feedback fixes signsgd and other gradient compression schemes.
\newblock In {\em International Conference on Machine Learning}, pages
  3252--3261, 2019.

\bibitem{grubic2018synchronous}
Demjan Grubic, Leo Tam, Dan Alistarh, and Ce~Zhang.
\newblock Synchronous multi-gpu deep learning with low-precision communication:
  An experimental study.
\newblock {\em Proceedings of the EDBT 2018}, 2018.

\bibitem{wu2018error}
Jiaxiang Wu, Weidong Huang, Junzhou Huang, and Tong Zhang.
\newblock Error compensated quantized sgd and its applications to large-scale
  distributed optimization.
\newblock In {\em International Conference on Machine Learning}, pages
  5321--5329, 2018.

\bibitem{stich2018sparsified}
Sebastian~Urban Stich, Jean-Baptiste Cordonnier, and Martin Jaggi.
\newblock Sparsified sgd with memory.
\newblock In {\em Advances in Neural Information Processing Systems}, pages
  4447--4458, 2018.

\bibitem{wen2017terngrad}
Wei Wen, Cong Xu, Feng Yan, Chunpeng Wu, Yandan Wang, Yiran Chen, and Hai Li.
\newblock Terngrad: Ternary gradients to reduce communication in distributed
  deep learning.
\newblock In {\em Advances in neural information processing systems}, pages
  1509--1519, 2017.

\bibitem{alistarh2017qsgd}
Dan Alistarh, Demjan Grubic, Jerry Li, Ryota Tomioka, and Milan Vojnovic.
\newblock Qsgd: Communication-efficient sgd via gradient quantization and
  encoding.
\newblock In {\em Advances in Neural Information Processing Systems}, pages
  1709--1720, 2017.

\bibitem{yu2019double}
Yue Yu, Jiaxiang Wu, and Longbo Huang.
\newblock Double quantization for communication-efficient distributed
  optimization.
\newblock In {\em Advances in Neural Information Processing Systems}, pages
  4440--4451, 2019.

\bibitem{lim20183lc}
Hyeontaek Lim, David~G Andersen, and Michael Kaminsky.
\newblock 3lc: Lightweight and effective traffic compression for distributed
  machine learning.
\newblock {\em arXiv preprint arXiv:1802.07389}, 2018.

\bibitem{elias1975universal}
Peter Elias.
\newblock Universal codeword sets and representations of the integers.
\newblock {\em IEEE transactions on information theory}, 21(2):194--203, 1975.

\bibitem{vogels2019powersgd}
Thijs Vogels, Sai~Praneeth Karimireddy, and Martin Jaggi.
\newblock Powersgd: Practical low-rank gradient compression for distributed
  optimization.
\newblock In {\em Advances in Neural Information Processing Systems}, pages
  14236--14245, 2019.

\bibitem{strom2015scalable}
Nikko Strom.
\newblock Scalable distributed dnn training using commodity gpu cloud
  computing.
\newblock In {\em Sixteenth Annual Conference of the International Speech
  Communication Association}, 2015.

\bibitem{dryden2016communication}
Nikoli Dryden, Tim Moon, Sam~Ade Jacobs, and Brian Van~Essen.
\newblock Communication quantization for data-parallel training of deep neural
  networks.
\newblock In {\em 2016 2nd Workshop on Machine Learning in HPC Environments
  (MLHPC)}, pages 1--8. IEEE, 2016.

\bibitem{chen2018adacomp}
Chia-Yu Chen, Jungwook Choi, Daniel Brand, Ankur Agrawal, Wei Zhang, and
  Kailash Gopalakrishnan.
\newblock Adacomp: Adaptive residual gradient compression for data-parallel
  distributed training.
\newblock In {\em Thirty-Second AAAI Conference on Artificial Intelligence},
  2018.

\bibitem{aji2017sparse}
Alham~Fikri Aji and Kenneth Heafield.
\newblock Sparse communication for distributed gradient descent.
\newblock In {\em Proceedings of the 2017 Conference on Empirical Methods in
  Natural Language Processing}, pages 440--445, 2017.

\bibitem{lin2017deep}
Yujun Lin, Song Han, Huizi Mao, Yu~Wang, and William~J Dally.
\newblock Deep gradient compression: Reducing the communication bandwidth for
  distributed training.
\newblock {\em arXiv preprint arXiv:1712.01887}, 2017.

\bibitem{sattler2019sparse}
Felix Sattler, Simon Wiedemann, Klaus-Robert M{\"u}ller, and Wojciech Samek.
\newblock Sparse binary compression: Towards distributed deep learning with
  minimal communication.
\newblock In {\em 2019 International Joint Conference on Neural Networks
  (IJCNN)}, pages 1--8. IEEE, 2019.

\bibitem{renggli2018sparcml}
C{\`e}dric Renggli, Dan Alistarh, Torsten Hoefler, and Mehdi Aghagolzadeh.
\newblock Sparcml: High-performance sparse communication for machine learning.
\newblock {\em arXiv preprint arXiv:1802.08021}, 2018.

\bibitem{alistarh2018convergence}
Dan Alistarh, Torsten Hoefler, Mikael Johansson, Nikola Konstantinov, Sarit
  Khirirat, and C{\'e}dric Renggli.
\newblock The convergence of sparsified gradient methods.
\newblock In {\em Advances in Neural Information Processing Systems}, pages
  5973--5983, 2018.

\bibitem{wangni2018gradient}
Jianqiao Wangni, Jialei Wang, Ji~Liu, and Tong Zhang.
\newblock Gradient sparsification for communication-efficient distributed
  optimization.
\newblock In {\em Advances in Neural Information Processing Systems}, pages
  1299--1309, 2018.

\bibitem{zhao2014kylix}
Huasha Zhao and John Canny.
\newblock Kylix: A sparse allreduce for commodity clusters.
\newblock In {\em 2014 43rd International Conference on Parallel Processing},
  pages 273--282. IEEE, 2014.

\bibitem{nguyen2019topology}
Truong~Thao Nguyen, Mohamed Wahib, and Ryousei Takano.
\newblock Topology-aware sparse allreduce for large-scale deep learning.
\newblock In {\em 2019 IEEE 38th International Performance Computing and
  Communications Conference (IPCCC)}, pages 1--8. IEEE, 2019.

\bibitem{fang2018redsync}
Jiarui Fang, Haohuan Fu, Guangwen Yang, and Cho-Jui Hsieh.
\newblock Redsync: Reducing synchronization traffic for distributed deep
  learning.
\newblock {\em arXiv preprint arXiv:1808.04357}, 2018.

\bibitem{yu2018gradiveq}
Mingchao Yu, Zhifeng Lin, Krishna Narra, Songze Li, Youjie Li, Nam~Sung Kim,
  Alexander Schwing, Murali Annavaram, and Salman Avestimehr.
\newblock Gradiveq: Vector quantization for bandwidth-efficient gradient
  aggregation in distributed cnn training.
\newblock In {\em Advances in Neural Information Processing Systems}, pages
  5123--5133, 2018.

\bibitem{wold1987principal}
Svante Wold, Kim Esbensen, and Paul Geladi.
\newblock Principal component analysis.
\newblock {\em Chemometrics and intelligent laboratory systems}, 2(1-3):37--52,
  1987.

\bibitem{wang2018atomo}
Hongyi Wang, Scott Sievert, Shengchao Liu, Zachary Charles, Dimitris
  Papailiopoulos, and Stephen Wright.
\newblock Atomo: Communication-efficient learning via atomic sparsification.
\newblock In {\em Advances in Neural Information Processing Systems}, pages
  9850--9861, 2018.

\bibitem{golub1971singular}
Gene~H Golub and Christian Reinsch.
\newblock Singular value decomposition and least squares solutions.
\newblock In {\em Linear Algebra}, pages 134--151. Springer, 1971.

\bibitem{zhang2017poseidon}
Hao Zhang, Zeyu Zheng, Shizhen Xu, Wei Dai, Qirong Ho, Xiaodan Liang, Zhiting
  Hu, Jinliang Wei, Pengtao Xie, and Eric~P Xing.
\newblock Poseidon: An efficient communication architecture for distributed
  deep learning on $\{$GPU$\}$ clusters.
\newblock In {\em 2017 $\{$USENIX$\}$ Annual Technical Conference
  ($\{$USENIX$\}$$\{$ATC$\}$ 17)}, pages 181--193, 2017.

\bibitem{shi2019mg}
Shaohuai Shi, Xiaowen Chu, and Bo~Li.
\newblock Mg-wfbp: Efficient data communication for distributed synchronous sgd
  algorithms.
\newblock In {\em IEEE INFOCOM 2019-IEEE Conference on Computer
  Communications}, pages 172--180. IEEE, 2019.

\bibitem{jayarajan2019priority}
Anand Jayarajan, Jinliang Wei, Garth Gibson, Alexandra Fedorova, and Gennady
  Pekhimenko.
\newblock Priority-based parameter propagation for distributed dnn training.
\newblock {\em arXiv preprint arXiv:1905.03960}, 2019.

\bibitem{hashemi2018tictac}
Sayed~Hadi Hashemi, Sangeetha~Abdu Jyothi, and Roy~H Campbell.
\newblock Tictac: Accelerating distributed deep learning with communication
  scheduling.
\newblock {\em arXiv preprint arXiv:1803.03288}, 2018.

\bibitem{peng2019generic}
Yanghua Peng, Yibo Zhu, Yangrui Chen, Yixin Bao, Bairen Yi, Chang Lan, Chuan
  Wu, and Chuanxiong Guo.
\newblock A generic communication scheduler for distributed dnn training
  acceleration.
\newblock In {\em Proceedings of the 27th ACM Symposium on Operating Systems
  Principles}, pages 16--29, 2019.

\bibitem{xu2020od}
Yemao Xu, Dezun Dong, Yawei Zhao, Weixia Xu, and Xiangke Liao.
\newblock Od-sgd: One-step delay stochastic gradient descent for distributed
  training.
\newblock {\em ACM Transactions on Architecture and Code Optimization (TACO)},
  17(4):1--26, 2020.

\bibitem{mai2015optimizing}
Luo Mai, Chuntao Hong, and Paolo Costa.
\newblock Optimizing network performance in distributed machine learning.
\newblock In {\em 7th $\{$USENIX$\}$ Workshop on Hot Topics in Cloud Computing
  (HotCloud 15)}, 2015.

\bibitem{gupta2016model}
Suyog Gupta, Wei Zhang, and Fei Wang.
\newblock Model accuracy and runtime tradeoff in distributed deep learning: A
  systematic study.
\newblock In {\em 2016 IEEE 16th International Conference on Data Mining
  (ICDM)}, pages 171--180. IEEE, 2016.

\bibitem{hsieh2017gaia}
Kevin Hsieh, Aaron Harlap, Nandita Vijaykumar, Dimitris Konomis, Gregory~R
  Ganger, Phillip~B Gibbons, and Onur Mutlu.
\newblock Gaia: Geo-distributed machine learning approaching $\{$LAN$\}$
  speeds.
\newblock In {\em 14th $\{$USENIX$\}$ Symposium on Networked Systems Design and
  Implementation ($\{$NSDI$\}$ 17)}, pages 629--647, 2017.

\bibitem{chilimbi2014project}
Trishul Chilimbi, Yutaka Suzue, Johnson Apacible, and Karthik Kalyanaraman.
\newblock Project adam: Building an efficient and scalable deep learning
  training system.
\newblock In {\em 11th $\{$USENIX$\}$ Symposium on Operating Systems Design and
  Implementation ($\{$OSDI$\}$ 14)}, pages 571--582, 2014.

\bibitem{cui2016geeps}
Henggang Cui, Hao Zhang, Gregory~R Ganger, Phillip~B Gibbons, and Eric~P Xing.
\newblock Geeps: Scalable deep learning on distributed gpus with a
  gpu-specialized parameter server.
\newblock In {\em Proceedings of the Eleventh European Conference on Computer
  Systems}, pages 1--16, 2016.

\bibitem{rabenseifner2004optimization}
Rolf Rabenseifner.
\newblock Optimization of collective reduction operations.
\newblock In {\em International Conference on Computational Science}, pages
  1--9. Springer, 2004.

\bibitem{sergeev2018horovod}
Alexander Sergeev and Mike Del~Balso.
\newblock Horovod: fast and easy distributed deep learning in tensorflow.
\newblock {\em arXiv preprint arXiv:1802.05799}, 2018.

\bibitem{baidu2017allreduce}
Baidu.
\newblock Baidu allreduce.
\newblock \url{https://github.com/baidu-research/baidu-allreduce}, 2017.

\bibitem{patarasuk2009bandwidth}
Pitch Patarasuk and Xin Yuan.
\newblock Bandwidth optimal all-reduce algorithms for clusters of workstations.
\newblock {\em Journal of Parallel and Distributed Computing}, 69(2):117--124,
  2009.

\bibitem{hintjens2013zeromq}
Pieter Hintjens.
\newblock {\em ZeroMQ: messaging for many applications}.
\newblock " O'Reilly Media, Inc.", 2013.

\bibitem{google2019grpc}
Google.
\newblock grpc.
\newblock \url{https://grpc.io}.

\bibitem{gropp2002mpich2}
William Gropp.
\newblock Mpich2: A new start for mpi implementations.
\newblock In {\em European Parallel Virtual Machine/Message Passing Interface
  Users’ Group Meeting}, pages 7--7. Springer, 2002.

\bibitem{facebook2017gloo}
FaceBook.
\newblock Gloo.
\newblock \url{https://github.com/facebookincubator/gloo}.

\bibitem{jeaugey2017nccl}
Sylvain Jeaugey.
\newblock Nccl 2.0.
\newblock {\em GTC}, 2017.

\bibitem{dryden2018aluminum}
Nikoli Dryden, Naoya Maruyama, Tim Moon, Tom Benson, Andy Yoo, Marc Snir, and
  Brian Van~Essen.
\newblock Aluminum: An asynchronous, gpu-aware communication library optimized
  for large-scale training of deep neural networks on hpc systems.
\newblock {\em 2018 IEEE/ACM Machine Learning in HPC Environments (MLHPC)},
  pages 1--13, 2018.

\bibitem{cho2019blueconnect}
Minsik Cho, Ulrich Finkler, and David Kung.
\newblock Blueconnect: Novel hierarchical all-reduce on multi-tired network for
  deep learning.
\newblock In {\em Proceedings of the Conference on Systems and Machine Learning
  (SysML)}, 2019.

\bibitem{mamidala2018mxnet}
Amith~R Mamidala, Georgios Kollias, Chris Ward, and Fausto Artico.
\newblock Mxnet-mpi: Embedding mpi parallelism in parameter server task model
  for scaling deep learning.
\newblock {\em arXiv preprint arXiv:1801.03855}, 2018.

\bibitem{vishnu2016distributed}
Abhinav Vishnu, Charles Siegel, and Jeffrey Daily.
\newblock Distributed tensorflow with mpi.
\newblock {\em arXiv preprint arXiv:1603.02339}, 2016.

\bibitem{awan2018optimized}
Ammar~Ahmad Awan, Ching-Hsiang Chu, Hari Subramoni, and Dhabaleswar~K Panda.
\newblock Optimized broadcast for deep learning workloads on dense-gpu
  infiniband clusters: Mpi or nccl?
\newblock In {\em Proceedings of the 25th European MPI Users' Group Meeting},
  pages 1--9, 2018.

\bibitem{awan2016efficient}
Ammar~Ahmad Awan, Khaled Hamidouche, Akshay Venkatesh, and Dhabaleswar~K Panda.
\newblock Efficient large message broadcast using nccl and cuda-aware mpi for
  deep learning.
\newblock In {\em Proceedings of the 23rd European MPI Users' Group Meeting},
  pages 15--22, 2016.

\bibitem{recio2007remote}
Renato Recio, Bernard Metzler, Paul Culley, Jeff Hilland, and Dave Garcia.
\newblock A remote direct memory access protocol specification.
\newblock Technical report, RFC 5040, October, 2007.

\bibitem{chu2006transmission}
Jerry Chu and Vivek Kashyap.
\newblock Transmission of ip over infiniband (ipoib).
\newblock Technical report, RFC 4391, April, 2006.

\bibitem{li2019improving}
Mingfan Li, Ke~Wen, Han Lin, Xu~Jin, Zheng Wu, Hong An, and Mengxian Chi.
\newblock Improving the performance of distributed mxnet with rdma.
\newblock {\em International Journal of Parallel Programming}, 47(3):467--480,
  2019.

\bibitem{liu2018optimizing}
Chang Liu, Jianwen Wei, Yi-Chao Wang, Minhua Wen, Simon See, and James Lin.
\newblock Optimizing deep learning frameworks incrementally to get linear
  speedup: a comparison between ipoib and rdma verbs.
\newblock In {\em 2018 IEEE 24th International Conference on Parallel and
  Distributed Systems (ICPADS)}, pages 473--480. IEEE, 2018.

\bibitem{xue2019fast}
Jilong Xue, Youshan Miao, Cheng Chen, Ming Wu, Lintao Zhang, and Lidong Zhou.
\newblock Fast distributed deep learning over rdma.
\newblock In {\em Proceedings of the Fourteenth EuroSys Conference 2019}, pages
  1--14, 2019.

\bibitem{biswas2018accelerating}
Rajarshi Biswas, Xiaoyi Lu, and Dhabaleswar~K Panda.
\newblock Accelerating tensorflow with adaptive rdma-based grpc.
\newblock In {\em 2018 IEEE 25th International Conference on High Performance
  Computing (HiPC)}, pages 2--11. IEEE, 2018.

\bibitem{jia2018improving}
Chengfan Jia, Junnan Liu, Xu~Jin, Han Lin, Hong An, Wenting Han, Zheng Wu, and
  Mengxian Chi.
\newblock Improving the performance of distributed tensorflow with rdma.
\newblock {\em International Journal of Parallel Programming}, 46(4):674--685,
  2018.

\bibitem{banerjee2016re}
Dip~Sankar Banerjee, Khaled Hamidouche, and Dhabaleswar~K Panda.
\newblock Re-designing cntk deep learning framework on modern gpu enabled
  clusters.
\newblock In {\em 2016 IEEE international conference on cloud computing
  technology and science (CloudCom)}, pages 144--151. IEEE, 2016.

\bibitem{chu2017efficient}
Ching-Hsiang Chu, Xiaoyi Lu, Ammar~A Awan, Hari Subramoni, Jahanzeb Hashmi,
  Bracy Elton, and Dhabaleswar~K Panda.
\newblock Efficient and scalable multi-source streaming broadcast on gpu
  clusters for deep learning.
\newblock In {\em 2017 46th International Conference on Parallel Processing
  (ICPP)}, pages 161--170. IEEE, 2017.

\bibitem{awan2017s}
Ammar~Ahmad Awan, Khaled Hamidouche, Jahanzeb~Maqbool Hashmi, and Dhabaleswar~K
  Panda.
\newblock S-caffe: Co-designing mpi runtimes and caffe for scalable deep
  learning on modern gpu clusters.
\newblock In {\em Proceedings of the 22nd ACM SIGPLAN Symposium on Principles
  and Practice of Parallel Programming}, pages 193--205, 2017.

\bibitem{ren2017irdma}
Yufei Ren, Xingbo Wu, Li~Zhang, Yandong Wang, Wei Zhang, Zijun Wang, Michel
  Hack, and Song Jiang.
\newblock irdma: Efficient use of rdma in distributed deep learning systems.
\newblock In {\em 2017 IEEE 19th International Conference on High Performance
  Computing and Communications; IEEE 15th International Conference on Smart
  City; IEEE 3rd International Conference on Data Science and Systems
  (HPCC/SmartCity/DSS)}, pages 231--238. IEEE, 2017.

\bibitem{xia2019rethinking}
Jiacheng Xia, Gaoxiong Zeng, Junxue Zhang, Weiyan Wang, Wei Bai, Junchen Jiang,
  and Kai Chen.
\newblock Rethinking transport layer design for distributed machine learning.
\newblock In {\em Proceedings of the 3rd Asia-Pacific Workshop on Networking
  2019}, pages 22--28. ACM, 2019.

\bibitem{wang2018bml}
Songtao Wang, Dan Li, Yang Cheng, Jinkun Geng, Yanshu Wang, Shuai Wang, Shu-Tao
  Xia, and Jianping Wu.
\newblock Bml: A high-performance, low-cost gradient synchronization algorithm
  for dml training.
\newblock In {\em Advances in Neural Information Processing Systems}, pages
  4238--4248, 2018.

\bibitem{guo2009bcube}
Chuanxiong Guo, Guohan Lu, Dan Li, Haitao Wu, Xuan Zhang, Yunfeng Shi, Chen
  Tian, Yongguang Zhang, and Songwu Lu.
\newblock Bcube: a high performance, server-centric network architecture for
  modular data centers.
\newblock In {\em Proceedings of the ACM SIGCOMM 2009 conference on Data
  communication}, pages 63--74, 2009.

\bibitem{al2008scalable}
Mohammad Al-Fares, Alexander Loukissas, and Amin Vahdat.
\newblock A scalable, commodity data center network architecture.
\newblock {\em ACM SIGCOMM computer communication review}, 38(4):63--74, 2008.

\bibitem{lecun1998gradient}
Yann LeCun, L{\'e}on Bottou, Yoshua Bengio, and Patrick Haffner.
\newblock Gradient-based learning applied to document recognition.
\newblock {\em Proceedings of the IEEE}, 86(11):2278--2324, 1998.

\bibitem{simonyan2014very}
Karen Simonyan and Andrew Zisserman.
\newblock Very deep convolutional networks for large-scale image recognition.
\newblock {\em arXiv preprint arXiv:1409.1556}, 2014.

\bibitem{guo2008dcell}
Chuanxiong Guo, Haitao Wu, Kun Tan, Lei Shi, Yongguang Zhang, and Songwu Lu.
\newblock Dcell: a scalable and fault-tolerant network structure for data
  centers.
\newblock In {\em Proceedings of the ACM SIGCOMM 2008 conference on Data
  communication}, pages 75--86, 2008.

\bibitem{yan2015performance}
Feng Yan, Olatunji Ruwase, Yuxiong He, and Trishul Chilimbi.
\newblock Performance modeling and scalability optimization of distributed deep
  learning systems.
\newblock In {\em Proceedings of the 21th ACM SIGKDD International Conference
  on Knowledge Discovery and Data Mining}, pages 1355--1364, 2015.

\bibitem{qi2017paleo}
Hang Qi, Evan~R Sparks, and Ameet Talwalkar.
\newblock Paleo: A performance model for deep neural networks.
\newblock (CONF), 2017.

\bibitem{abadi2016tensorflow}
Mart{\'\i}n Abadi, Paul Barham, Jianmin Chen, Zhifeng Chen, Andy Davis, Jeffrey
  Dean, Matthieu Devin, Sanjay Ghemawat, Geoffrey Irving, Michael Isard, et~al.
\newblock Tensorflow: A system for large-scale machine learning.
\newblock In {\em 12th $\{$USENIX$\}$ Symposium on Operating Systems Design and
  Implementation ($\{$OSDI$\}$ 16)}, pages 265--283, 2016.

\bibitem{chen2015mxnet}
Tianqi Chen, Mu~Li, Yutian Li, Min Lin, Naiyan Wang, Minjie Wang, Tianjun Xiao,
  Bing Xu, Chiyuan Zhang, and Zheng Zhang.
\newblock Mxnet: A flexible and efficient machine learning library for
  heterogeneous distributed systems.
\newblock {\em arXiv preprint arXiv:1512.01274}, 2015.

\bibitem{xu2019sketchdlc}
Yemao Xu, Dezun Dong, Weixia Xu, and Xiangke Liao.
\newblock Sketchdlc: A sketch on distributed deep learning communication via
  trace capturing.
\newblock {\em ACM Transactions on Architecture and Code Optimization (TACO)},
  16(2):1--26, 2019.

\end{thebibliography}
\end{document}